%% file: main.tex
\documentclass[conference]{IEEEtran}

\usepackage{tikz}
\newcommand*\circled[1]{\tikz[baseline=(char.base)]{
            \node[shape=circle,draw,inner sep=2pt] (char) {#1};}}

\usepackage{amsmath,amssymb,amsfonts}
\usepackage{algorithmic}
\usepackage{graphicx}
\usepackage[most]{tcolorbox}
\usepackage{textcomp}
\usepackage{bmpsize}
\usepackage{xcolor}
\usepackage{lipsum}
\usepackage[braket, qm]{qcircuit}
\usepackage{filecontents}
\usepackage{algorithm}
\usepackage{caption}
\usepackage{subcaption}
\usepackage{adjustbox}
\usepackage{setspace}
\usepackage{tabularx}
\usepackage{balance}
\usepackage{pifont}
\usepackage{bbding}
\usepackage{multicol,multirow}
\usepackage{color}
\usepackage{soul}
\tcbset{
  mytextbox/.style={
    colback=blue!10!white,      
    colframe=blue!30!white, 
    fonttitle=\bfseries,
    coltitle=black,
    title={Key Takeaways:},
    boxrule=0.8pt,
    arc=4pt,
    left=6pt, right=6pt, top=6pt, bottom=6pt
  }
}

\begin{document}

\title{QubitHammer: Remotely Inducing Qubit State Change on Superconducting Quantum Computers}


\author{\IEEEauthorblockN{Yizhuo Tan}
\IEEEauthorblockA{
\textit{Yale University}\\
New Haven, USA \\
yizhuo.tan@yale.edu}
\and
\IEEEauthorblockN{Navnil Choudhury}
\IEEEauthorblockA{
\textit{University of Texas at Dallas}\\
Richardson, USA \\
Navnil.Choudhury@utdallas.edu}
\and
\IEEEauthorblockN{Kanad Basu}
\IEEEauthorblockA{
\textit{University of Texas at Dallas}\\
Richardson, USA \\
Kanad.Basu@utdallas.edu}
\and
\IEEEauthorblockN{Jakub Szefer}
\IEEEauthorblockA{
\textit{Northwestern University}\\
Evanston, USA \\
jakub.szefer@northwestern.edu}
}

\maketitle

\begin{abstract}
\input{sections/abstract}
\end{abstract}




\pagestyle{plain}

\input{sections/introduction}

\input{sections/background}

\input{sections/threat_model}

\input{sections/attack_methods}

\input{sections/attack_types}

\input{sections/exp_setup}

\input{sections/ibm_exp_results}

\input{sections/rigetti_exp_results}

\input{sections/evaluation_summary}

\input{sections/defenses}

\input{sections/related_work}

\input{sections/conclusion}

\clearpage

\input{sections/ethical_consideration}

\bibliographystyle{IEEEtran}
\bibliography{main}

\end{document}

%% file: sections/abstract.tex
To address the rapidly growing demand for cloud-based quantum computing, various researchers are proposing shifting from the existing single-tenant model to a multi-tenant model that expands resource utilization and improves accessibility. However, while multi-tenancy enables multiple users to access the same quantum computer, it introduces potential for security and reliability vulnerabilities. It therefore becomes important to investigate these vulnerabilities, especially considering realistic attackers who operate without elevated privileges relative to an ordinary users. To address this research need, this paper presents and evaluates QubitHammer, the first attack to demonstrate that an adversary can remotely induce unauthorized changes to a victim's quantum circuit's qubit’s state within a multi-tenant model by using custom qubit control pulses that are generated within constraints of the public interfaces and without elevated privileges. Through extensive evaluation on real-world superconducting devices from IBM and Rigetti, this work demonstrates that QubitHammer allows an adversary to significantly change the output distribution of a victim quantum circuit. In the experimentation, variational distance is used to evaluate the magnitude of the changes, and variational distance as high as \textit{0.938} is observed. Cross-platform analysis of QubitHammer on the number of quantum computing devices exposes a fundamental susceptibility in superconducting hardware. Futher, QubitHammer was also found to evade all currently proposed defenses aimed at ensuring reliable execution in multi-tenant superconducting quantum systems.

%% file: sections/introduction.tex
\section{Introduction}

In recent years, there has been a significant surge in the development and deployment of Noisy Intermediate-Scale Quantum (NISQ) computers, typically in the form of quantum processing units (QPUs)~\cite{Preskill2018quantumcomputingin}. 
These QPUs are constructed using qubits, which serve as the fundamental units of quantum information, interconnected by quantum gates and couplers that enable entanglement and coherent quantum operations across the qubit array.
Popular technologies for building QPUs include superconducting qubits~\cite{ibmquantum}, trapped ions~\cite{quantinuum}, neutral atoms~\cite{quera}, silicon spin qubits~\cite{neyens2024probing}, photons~\cite{zhong2020quantum}, and diamond NV centers~\cite{QTI}. Quantum systems are poised to revolutionize computation, achieving complex tasks that would take classical computers exponentially longer in comparison. 

Access to these quantum computers can be obtained either by purchasing an individual system or, more commonly, through cloud-based platforms. Cloud services like Microsoft Azure, qBraid, Amazon Braket, IBM Quantum, and others enable users to run jobs on larger quantum processors~\cite{prateek2023quantum, qbraid2024, gonzalez2021cloud}. 
With increased accessibility, applications of quantum computers in several fields are gaining traction, including machine learning, cryptography, and molecular chemistry~\cite{chemestry, easttom2022quantum, sajwan2019challenges}. As more applications of quantum computing become apparent, the demand for quantum computing resources continues to surge.
However, existing quantum computers operate in a single-tenant mode, where only one user has access to a QPU at any time. As quantum computers grow in size, ingle-tenancy leads to low throughput and the underutilization of valuable quantum~resources.

\begin{figure*}[t]
\centering
 \begin{subfigure}[b]{0.32\textwidth}
 \includegraphics[width=\textwidth]{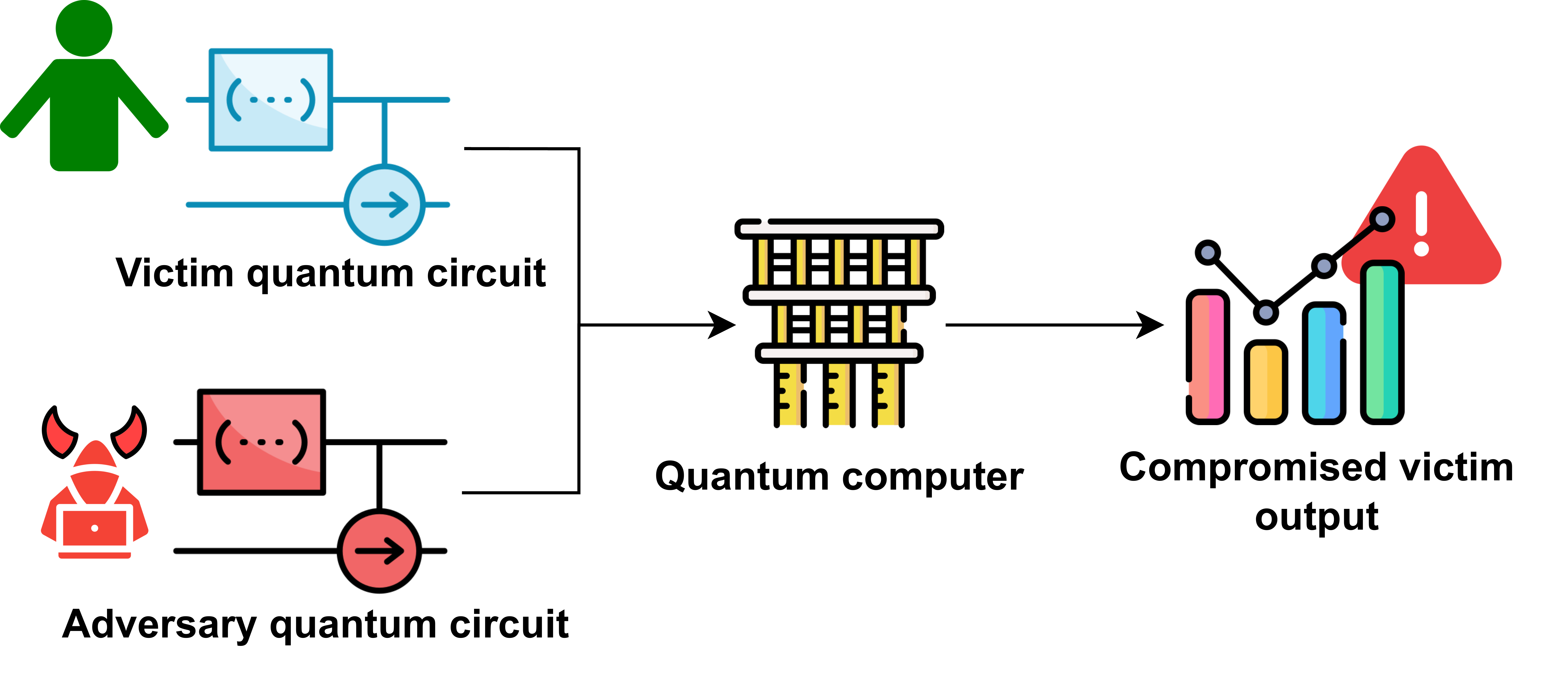}
 \caption{\small Sharing of the same hardware resulting in compromised victim computation.}
 \label{subfig:threat}
 \end{subfigure}
 \hspace{0.1cm}
 \begin{subfigure}[b]{0.32\textwidth}
 \centering
 \includegraphics[width=0.8\textwidth]{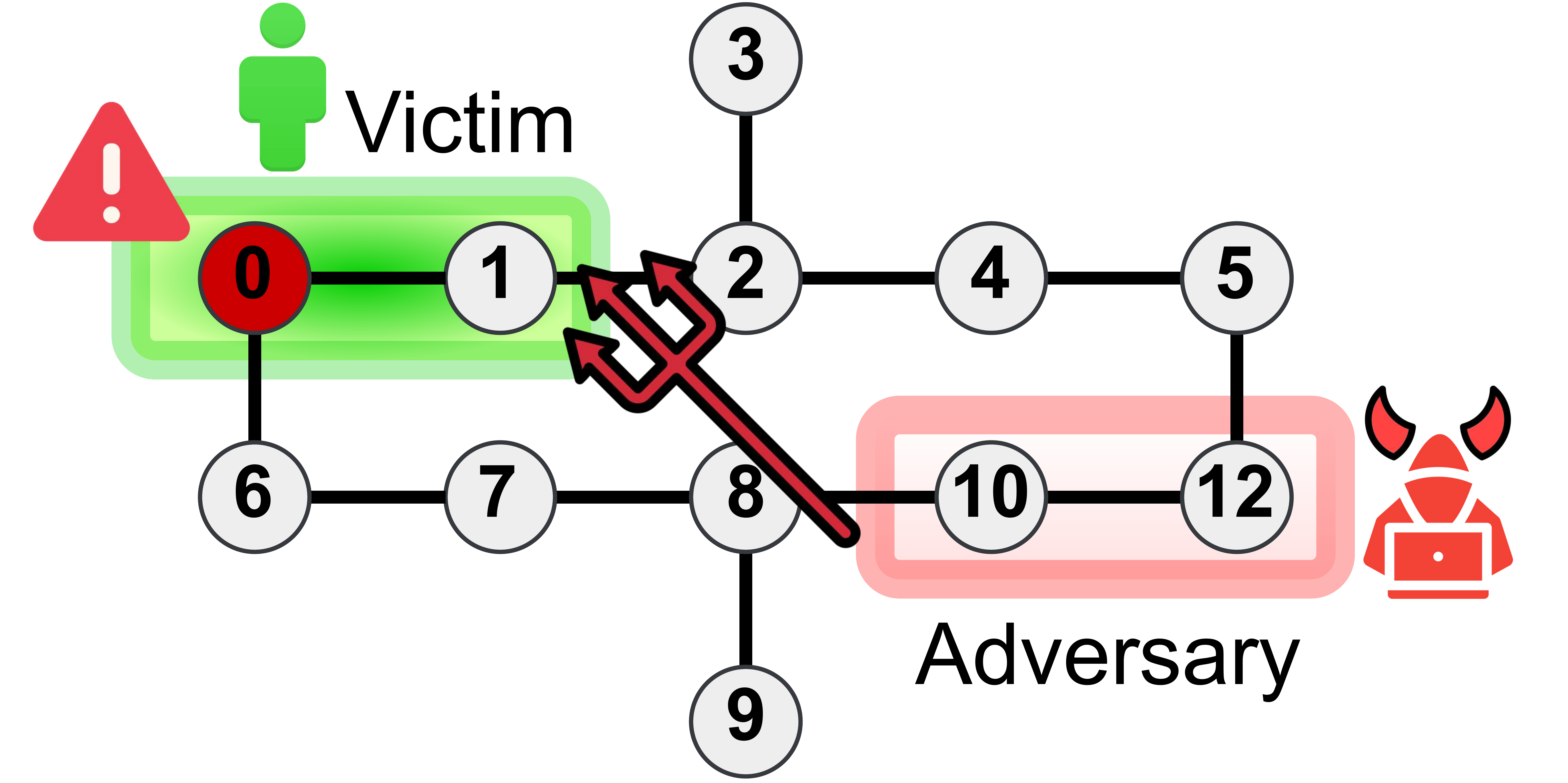}
 \caption{\small An example of a 12-qubit quantum computer where qubits {0, 1} are used by a victim.}
 \label{subfig:victim}
 \end{subfigure}
 \hspace{0.1cm}
 \begin{subfigure}[b]{0.32\textwidth}
 \centering
 \includegraphics[width=0.8\textwidth]{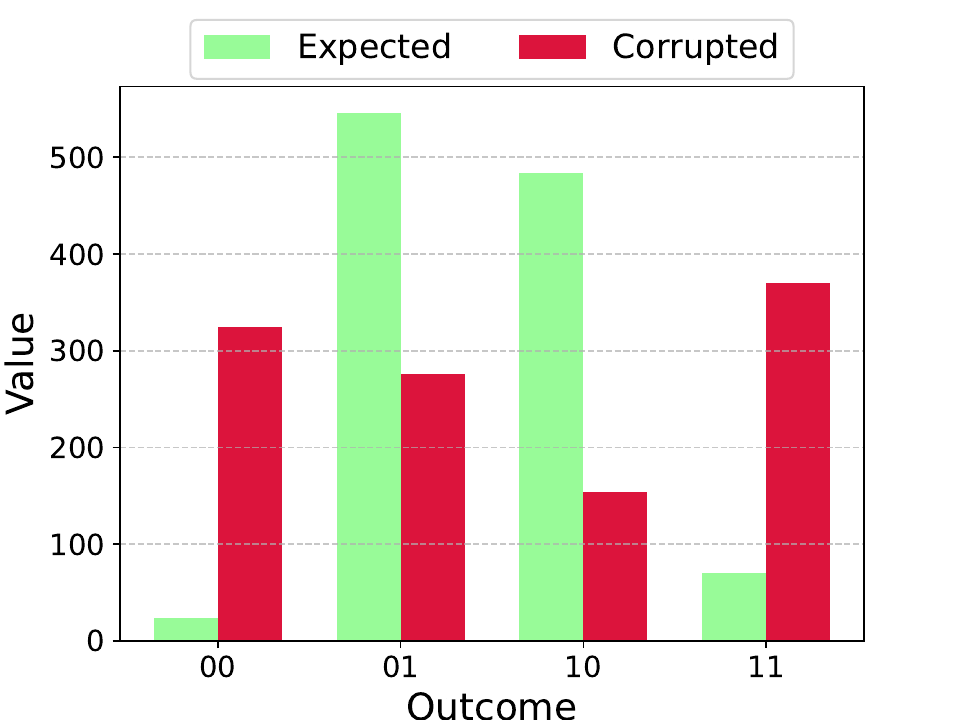}
 \centering
 \caption{\small Adversary using qubits 10, 12 to launch QubitHammer, flipping state of qubit 0.}
 \label{subfig:motivation_res}
 \end{subfigure}
\vspace{-4mm}
\caption{\small Overview of proposed QubitHammer attack. Using designed attack pulses, an adversary can flip the state of a victim qubit, compromising the reliability of computation.}
\label{fig:atk_intro}
\end{figure*}

To address this inefficiency and improve the utilization of limited quantum resources, researchers have explored methods for enabling multi-tenancy in quantum computers, where multiple users can execute their programs simultaneously on the same quantum computer~\cite{das2019case}~\cite{liu2021qucloud}. By partitioning the quantum computer into separate segments that execute quantum circuits in parallel on non-overlapping sets of qubits, they aim to maximize the utilization of these cloud-based quantum systems. Currently, three promising approaches exist for robust multi-tenancy~\cite{das2019case}. The Fair and Reliable Partitioning (FRP) algorithm allocates reliable qubits to each program based on machine calibration error data. To address the impact of qubit measurement operations on other simultaneously executed operations of varying lengths~\cite{das2019case}, the Delayed Instruction Scheduling (DIS) policy reschedules the start times of these programs~\cite{das2019case}. Lastly, when multi-tenancy significantly affects program reliability, quantum hardware providers can switch to isolated execution mode using an Adaptive Multi-Programming (AMP) design~\cite{das2019case}.

While the proposed multi-tenancy enhances the utilization of quantum computers significantly, it also introduces new challenges when multiple users can execute their programs on the same quantum computer concurrently.
A comparison of recent studies demonstrating security issues in multi-tenant quantum computing systems is provided in Table~\ref{tab:multi_tenancy}. These studies explore power side-channels, timing-side-channels, and crosstalk-based side-channel attacks to disrupt computation or steal information from quantum computers. 
Our work focuses on remotely corrupting victim qubit states using attack pulses, including inducing qubit flips. Unlike prior efforts that were limited to adjacent qubits (with minimal separation in the quantum processor topology), this is the first demonstration that an adversary can influence non-adjacent and physically distant victim qubits on a superconducting quantum chip, without requiring the prominent presence of two-qubit gates to induce crosstalk.

\textcolor{black}{
Our  attack is named QubitHammer, and it realizes a novel reliability attack strategy targeting superconducting quantum hardware, where an adversary can remotely induce errors in a co-located victim's computation. An overview of this attack is depicted in Figure~\ref{fig:atk_intro}. The figure shows that an adversary can design and deploy a quantum program that compromises the integrity and reliability of victim circuit execution on the same quantum computer. For example, a victim circuit on qubits 0 and 1 is co-located with an adversary's program on distant qubits 10 and 12 on a 12-qubit quantum computer. In this scenario, QubitHammer enables the adversarial user to remotely corrupt the victim's computation by modifying the state of qubit 0, demonstrating a non-local security vulnerability.
The impact of QubitHammer in this scenario is shown in Figure~\ref{subfig:motivation_res}, where the x-axis denotes the possible outcomes, while the y-axis denotes the value of fidelity.
The QubitHammer attack disrupts the computation by skewing the output distribution away from the expected '00' and '01' states and toward the erroneous state '11', which becomes the dominant outcome.}
To understand the significance of this attack in the real world, consider a scenario where the victim has deployed a Variational Quantum Eigensolver (VQE) for a portfolio optimization algorithm~\cite{elsokkary2017financial}~\cite{rebentrost2018quantum}. The successful deployment of QubitHammer by the adversary co-located on the same quantum computing chip would result in the unauthorized flipping of the states of one or more qubits, resulting in completely inaccurate risk evaluations or flawed investment strategies based on a compromised output. Such a scenario would ultimately result in significant financial losses for institutions relying on quantum computing for high-stakes decision-making.

\begin{table}[b]
\centering
\caption{\small Overview of recent studies on security challenges in multi-tenant quantum computers.}
\resizebox{\linewidth}{!}{%
\begin{tabular}{c|c}
\hline
\textbf{Related Work}      &    \textbf{Attack}\\ \hline
Deshpande \textit{et al}. (CCS 2022)~\cite{deshpande2022towards} & 
Adjacent Qubit crosstalk\\ \hline
Erata \textit{et al}. (CCS 2023)~\cite{erata2024quantum} & 
Controller Side-Channel \\ \hline
Upadhyay \textit{et al}. (VLSID 2024)~\cite{upadhyay2024stealthy} & 
Adversarial SWAP injection  \\ \hline
Choudhury \textit{et al}. (NDSS 2025)~\cite{choudhury2024crosstalkinducedchannelthreatsmultitenant}  &
Victim circuit reconstruction  \\ \hline
\end{tabular}%
}
\label{tab:multi_tenancy}
\end{table}

To understand impact of QubitHammer,
we present the first in-depth study of various attack variants aimed at disrupting quantum operations in a multi-tenant environment. 
Our findings reveal that the adversary can cause significant disruption to the victim's quantum circuit execution by deploying specially designed attack pulses on their own qubits, which cause far-away victim qubits to be disturbed. 
The ideas and impact of QubitHammer is demonstrated on both IBM and Rigetti quantum computers. Further, prior defenses such as keeping qubits idle~\cite{ash2020analysis} to separate adversary and victim do not work, as we can induce qubit flipping in far away qubits. Based on the new understanding of the vulnerabilities, safe and secure multi-tenancy for quantum computers can be designed.

\subsection{Contributions}

\begin{itemize}
    \setlength\itemsep{0em}
    \item In this work, we present QubitHammer, a novel class of indirect-access, pulse-level adversarial attacks, capable of remotely manipulating the quantum state of a victim qubit and thereby introducing a novel reliability challenge for multi-tenant superconducting quantum processors. 
    \item We illustrate a novel attack pulse generation method, where the adversary leverages real-time public-access data of victim qubits to successfully disrupt computation of a victim.
    \item We explore the generation of different attack pulse configurations available to an adversary, and their impact on different victim qubits across well-defined attack scenarios, each reflecting realistic qubit allocation schemes on today’s quantum hardware.
    \item We evaluate QubitHammer under realistic threat assumptions across different benchmarks on both IBM and Rigetti superconducting platforms, achieving variational distances as high as \textit{0.938}, which highlights cross-architecture applicability of the attack and its severe impact on computational reliability.
    \item Finally, we systematically evaluate QubitHammer’s effectiveness against existing countermeasures in real quantum hardware, achieving variational distances as high as \textit{0.837}, meaning existing defenses do not effectively mitigate such new attack.
\end{itemize}

%% file: sections/background.tex
\section{Background}

This section presents background on quantum computation, as well as on NISQ quantum computers and today's cloud-based deployments of quantum computers.

\subsection{Quantum Computing Principles}

Unlike the classical bit, which can be either 0 or 1, a quantum bit (qubit for short) can be a linear combination of two basis states $\ket{0}$ and $\ket{1}$. A qubit state $\ket{\psi}$ can be represented as:
\begin{align}
    \ket{\psi} = \alpha\ket{0}+\beta\ket{1}
\end{align}
where $\alpha$ and $\beta$ are complex numbers which satisfy $|\alpha|^2 + |\beta|^2 = 1$.
More generally, \textit{n}-qubit state $\ket{\phi}$ can be expressed as follow:
\begin{align}
    \ket{\phi} = \sum_{i=0}^{2^n-1}\alpha_i\ket{i}
\end{align}
where $\ket{i}$ is one of $2^n$ basis states from $\ket{0...0}$ to $\ket{1...1}$, $\alpha_i$ satisfies $\sum_{i=0}^{2^n-1}|\alpha_i|^2 = 1$. 

This superposition principle allows quantum computers to explore many states simultaneously and provides massive potential computational power. The \textit{n}-qubit quantum states can also be represented using n-dimensional vectors. For example, $\ket{00...0} = [1,0,...,0]^T$ and $\ket{1,...,1,1} = [0,...,0,1]^T$. As a result, the unitary quantum gate $U$, which satisfies $UU^T = U^TU = I$, operating qubits like $U\ket{\phi}$ can be represented by $2^n\times2^n$ matrix. Here are some examples of frequently used gates:
{\footnotesize
\begin{equation*}
    X = 
    \begin{bmatrix}
        0 & 1 \\
        1 & 0 \\
    \end{bmatrix}, 
    SX = {1\over2}
    \begin{bmatrix}
        1+i & 1-i \\
        1-i & 1+i \\
    \end{bmatrix}, 
    CX = 
    \begin{bmatrix}
        1 & 0 & 0 & 0 \\
        0 & 0 & 0 & 1 \\
        0 & 0 & 1 & 0 \\
        0 & 1 & 0 & 0 \\
    \end{bmatrix}
\end{equation*}}
Most quantum gates, such as the {\tt Hadamard} gate, need to be decomposed into some basis gates before submitting to the real quantum computer hardware.
\label{subsec:quantum_computing_principle}

\begin{figure}[t]
\begin{subfigure}[b]{0.58\linewidth}
\centering
\includegraphics[width=\textwidth]{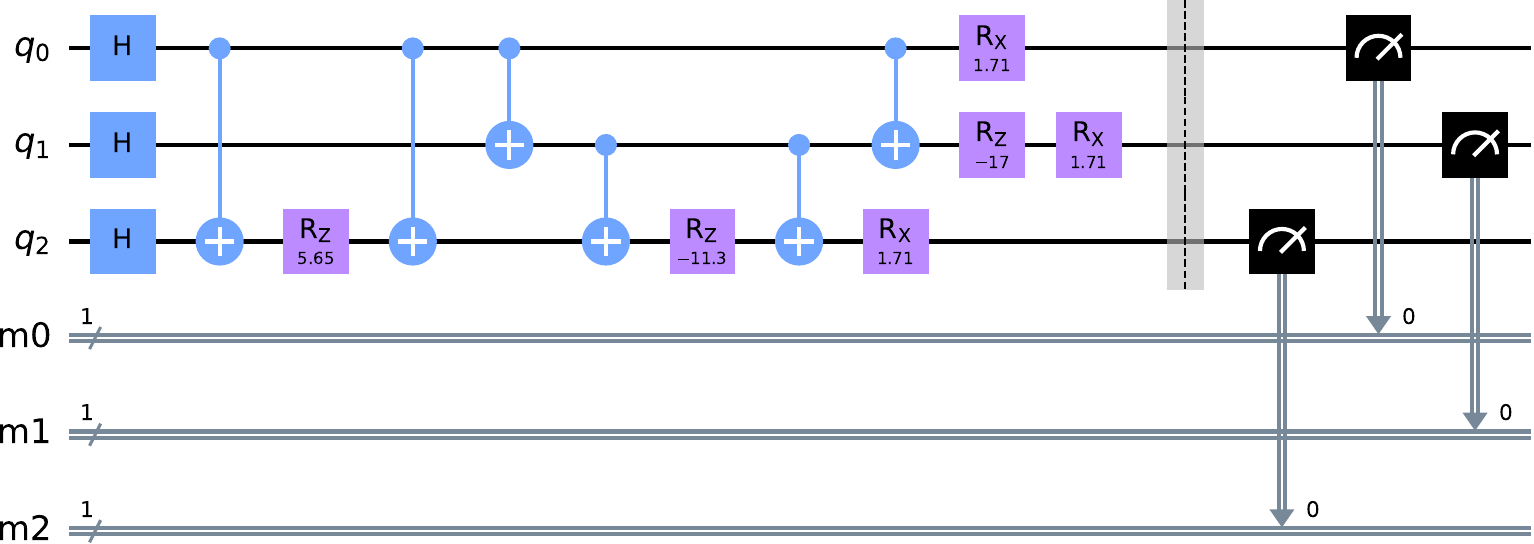}
\caption{\small An example quantum circuit with three qubits.}
\label{subfig:qaoa_ex}
\end{subfigure}
\hfill
\begin{subfigure}[b]{0.38\linewidth}
\centering
\includegraphics[width=\textwidth]{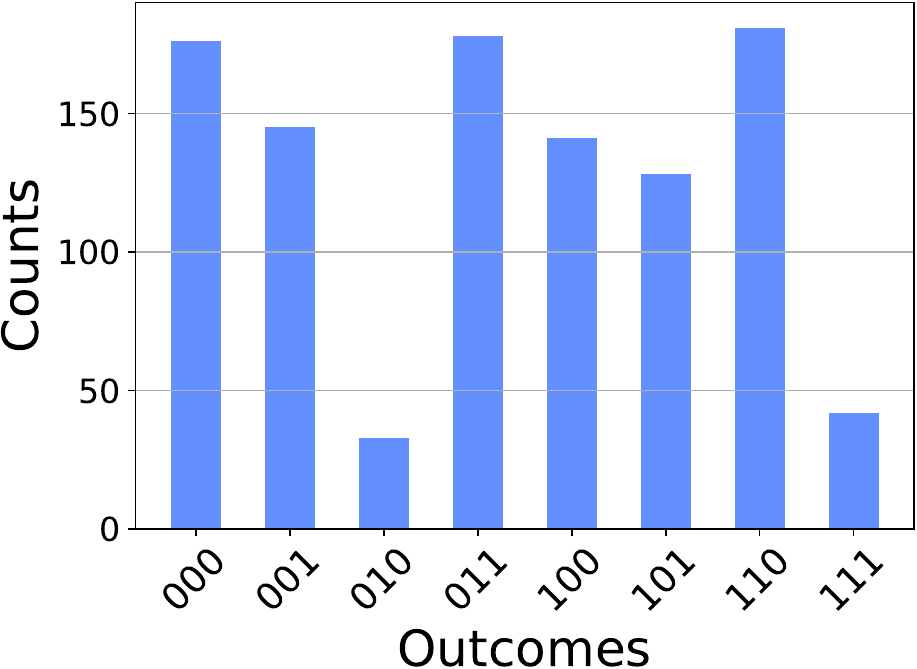}
\caption{\small Example output obtained from the circuit.}
\label{subfig:qaoa_res_ex}
\end{subfigure}
\caption{\small Example of quantum circuit and it's output probabilities: a $3$ qubit circuit will have $2^3=8$ output states each with some probability. The classical ``result'' of the computation is determined from the output count probabilities.}
\label{fig:ex_ckt_exec}
\end{figure}

\subsection{Quantum Circuit}

A quantum circuit comprises a sequence of quantum gates (Section~\ref{subsec:quantum_computing_principle}) applied to qubits and followed by measurement. After constructing a logic-level circuit using a development kit such as Qiskit~\cite{qiskit2024}, it must be transpiled into hardware-specific instructions. Figure~\ref{subfig:qaoa_ex} illustrates an example with three qubits and corresponding gates. Transpilation maps the circuit to basis gates compatible with the target hardware topology. Superconducting qubits are typically controlled via microwave pulses, which are generated during the scheduling stage~\cite{qiskit2024} that converts gate-level to pulse-level instructions. Once all gates are applied, the qubits are measured. An example of resulting measurement probabilities is shown in Figure~\ref{subfig:qaoa_res_ex}. A quantum circuit with $n$ qubits can yield $2^n$ possible outcomes. However, due to the presence of noise in NISQ devices, each quantum circuit needs to be executed thousands of times ({each execution is called a \em shot}), for reliable probability estimations. The final output is derived from these probabilities, and modifying them changes the output, adversely impacting the computation reliability.

\subsection{Pulse-Level Control}

Superconducting quantum computers use microwave pulses (defined by envelope, frequency, and phase) to control qubits~\cite{qiskit2024}. The envelope shapes the signal via an arbitrary waveform generator, while frequency and phase modulate it. These combined signals drive qubit operations. Since each qubit has a unique frequency, identical operations require qubit-specific pulse parameters. Before execution, a gate-level circuit (e.g., Figure~\ref{subfig:qaoa_ex}) is compiled into a pulse-level form by decomposing gates into native pulses supported by the hardware. IBM~\cite{qiskit2024} and Rigetti provide predefined, calibrated native pulses and also support custom pulse-level programming via low-level interfaces.
\textit{It is important to note that pulse-level access to a quantum computer is not a special privilege, but a standard capability offered to any normal user through the cloud provider's API.
}

\subsection{Quantum-as-a-Service}

Similar to Platform-as-a-Service (PaaS) or Infrastructure-as-a-Service (IaaS), Quantum-as-a-Service (QaaS)~\cite{qaas} refers to cloud-based delivery of quantum technologies, quantum computing services and quantum computing solutions. without needing to own or maintain the hardware. 
Superconducting quantum computers from IBM~\cite{ibmquantum}, Rigetti~\cite{rigetti}, or IQM are all available through cloud-based platforms today. 
Users typically submit their quantum circuits to the cloud, where they enter a queue awaiting execution. Before execution, each circuit is transpiled for the target hardware. This process selects an appropriate subset of physical qubits, breaks down high‑level gates into the device’s native instruction set, and then schedules these instructions to maximize overall fidelity.

\subsection{Multi-tenant Quantum Computers}
The growing number of quantum computer users has caused a demand-supply imbalance, resulting in substantial wait times for accessing quantum computing resources~\cite{ravi2021quantum}. Hence, a multi-tenant quantum computing system is preferred in terms of cost efficiency, resource utilization, and accessibility. Through mapping multiple quantum programs onto a single quantum hardware simultaneously, multi-tenant Quantum computers can be shared by multiple users at the same time~\cite{das2019case}~\cite{liu2021qucloud}. This is crucial for making quantum computing more accessible and efficient, especially given the high cost of quantum hardware and long waiting time for queues. However, multi-tenant environments give rise to multiple security and reliability issues that have been studied.

%% file: sections/threat_model.tex
\section{Threat Model} \label{sec:threat}

Our threat model assumes minimal adversarial privileges, resulting in a realistic attack that can be deployed in today's quantum computing environments with only user-level, pulse-level access, and enabled multi-tenancy. The attack name, QubitHammer, is inspired from the well-studied RowHammer attack in classical computers~\cite{kim2014flipping}. Although physical realizations of quantum and classical computers are different, the wide threat of RowHammer in classical computers motivates the need to study QubitHammer, which poses a similar threat.

\begin{figure}[t]
    \centering
    \includegraphics[width=0.9\linewidth]{figures/multi_tenancy_ibm.pdf}
    \vspace{-3mm}
    \caption{\small A multi-tenant IBM quantum computer with four users, where a malicious user can induce unauthorized changes to a victim's computation, modifying their qubit states resulting in the wrong outcome. The topology is based on the \textit{IBM Eagle} QPUs.}
    \label{fig:threat_model_ibm}
\end{figure}

\textbf{Assumptions about Adversary's Access:}
We assume that the adversary can execute their quantum circuits on the same quantum computer as the victim, albeit on separate, non-overlapping sets of qubits. An example scenario is shown in Figure~\ref{fig:threat_model_ibm} 
where, e.g., four users (three victims and one adversary) are concurrently executing their quantum programs on the same quantum computer that enables multi-tenancy. 
\textit{Note that our threat model does not assume co-location of victim and adversary circuits on physically adjacent qubits; the adversary and victim can be very far part, unlike for other crosstalk attacks.}


\textbf{Assumptions on Adversary's Objective}
%
We assume the adversary’s goal is to disrupt quantum computation by inducing unauthorized state changes in victim qubits, leading to denial-of-service or incorrect outputs. This is achieved by exploiting crosstalk and related effects between adversary-controlled qubits and nearby victim qubits.

\textbf{Assumptions on Adversary's Capabilities}
\label{subsec:assumptions}
The adversary is assumed to have four minimal, realistic capabilities: (1) co-location with the victim on the same multi-tenant quantum computer; (2) pulse-level access, which is standard for users; (3) knowledge of the victim’s execution timing. This is a reasonable assumption, since the repeated execution (shots) of any quantum program makes temporal overlap with an adversary likely; 
and (4) control over their own qubit states, permitted under typical logical isolation.


%% file: sections/attack_methods.tex
\section{QubitHammer Attack}
\label{sec:attack_method}

In this section, we delineate our proposed QubitHammer attack, as illustrated in Figure~\ref{fig:atk_method}. 
As described in the figure, setting up the QubitHammer attack involves a two-step approach, namely: \circled{1} designing attack pulses, and \circled{2} using them to drive adversarial qubits (resulting in crosstalk that affects the far-away victim's qubits). It is important to note that the adversary can use single or multiple attack pulses on their qubits. 

\subsection{Attack Pulse Design}
\label{subsec:pulse_design}
To deploy QubitHammer, we begin by designing custom attack pulses that will subsequently drive the adversarial qubits. We follow a two step process, outlined as follows: 

\subsubsection{Frequency Sweep} 
\begin{figure}[b]
 \centering
 \begin{subfigure}[b]{0.45\linewidth}
 \includegraphics[width=\textwidth]{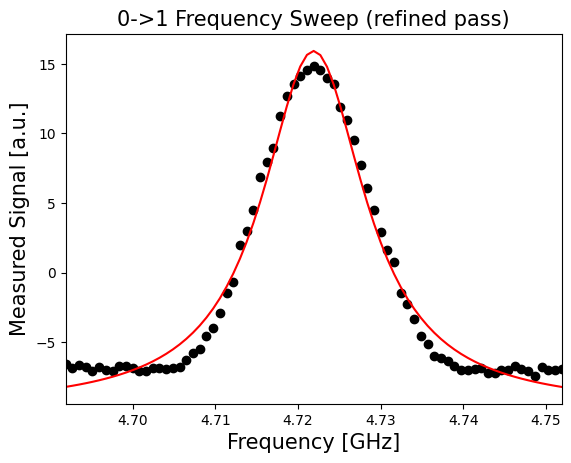}
 \caption{\small Frequency Sweep on a target qubit for \texttt{X} gate.}
 \label{subfig:frequency_sweep}
 \end{subfigure}
 \begin{subfigure}[b]{0.45\linewidth}
 \includegraphics[width=\textwidth]{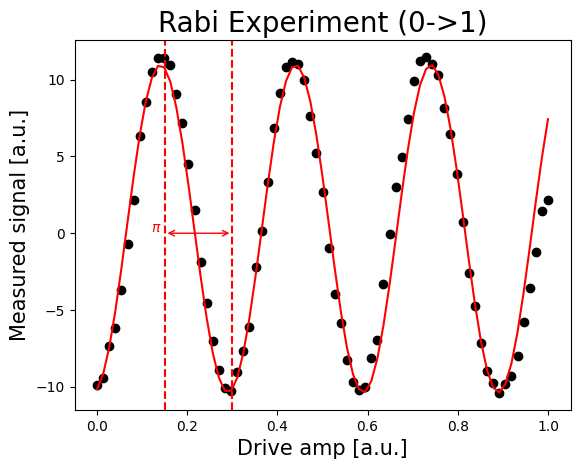}
 \caption{\small Rabi Oscillaton Experiment on a target qubit for \texttt{X} gate.}
 \label{subfig:rabi}
 \end{subfigure}
 \caption{\small Demonstration of the two-step calibration experiment on \textit{IBM\_brisbane} processor.}
 \label{fig:pulse_design}
\end{figure}

First, we conduct a frequency spectroscopy to determine the target qubit's true desired transition frequency ($f_0$). We begin by generating a sequence of identical Gaussian-shaped pulses with fixed amplitude and duration on a target qubit's drive channel. These pulses are applied across a detuning window centered on the qubit’s nominal transition frequency. Within this window, the drive frequency is stepped in sub-MHz increments to provide enough spectral resolution for detecting small drifts while keeping the overall sweep duration practical. For example, when calibrating an X gate, the detuning window is centered on the qubit’s $\ket{0} \rightarrow \ket{1}$ transition, as shown in Figure~\ref{subfig:frequency_sweep}. At each frequency step, we average the readout resonator’s in-phase and quadrature voltages over many shots to reduce quantum and amplifier noise. When the drive hits the qubit’s true resonance, the qubit absorbs energy and a clear peak appears in the averaged signal. Fitting that response with a Lorentzian curve then reveals the qubit’s exact transition frequency.

\begin{figure*}[t]
    \centering
    \includegraphics[width=0.8\linewidth]{figures/threat_model_n4.pdf}
    \vspace{-1mm}
    \caption{\small Overview of our proposed QubitHammer attack.}
    \label{fig:atk_method}
\end{figure*}

\subsubsection{Rabi Oscillation Experiment (ROE)} Following the frequency sweep, we perform an ROE to determine the duration (amplitude) of the pulse ($T_\theta$) that produces our desired rotation, when driving the target qubit at $f_0$. To this end, we being by fixing the microwave drive to the resonance frequency identified during spectroscopy ($f_0$) and apply a series of identical Gaussian-shaped pulses whose durations are systematically varied. After each pulse, we measure the qubit’s excited-state probability over many shots to average out quantum and amplifier noise. As the pulse length increases, the qubit undergoes coherent Rabi oscillations between its basis states, producing a sinusoidal variation in the measured excitation probability. Plotting this probability versus pulse duration reveals distinct oscillations whose first peak corresponds to a desired rotation angle $\theta$. By fitting these oscillations to a (damped) sinusoidal model, we extract the exact pulse duration needed to perform a high-fidelity $\theta$ rotation. For example, for an X gate, the first oscillation peak between its basis states $\ket{0}$ and $\ket{1}$ gives the $\pi$ pulse duration needed for a full bit-flip. This is depicted in Figure~\ref{subfig:rabi}.

\subsubsection{Attack Pulse Construction}
Building on the transition frequency \(f_0\) and pulse duration \(T_\theta\) obtained from the preceding spectroscopy and Rabi experiments, we synthesize custom Gaussian attack pulses on the adversary’s drive channel. Each pulse is configured with the extracted \(f_0\) and \(T_\theta\) to enact the desired rotation angle \(\theta\) on the victim qubit. This on-the-fly synthesis locks the attack pulses to the qubit’s instantaneous operating conditions—automatically compensating for any drift in frequency or amplitude—and thereby maximizes the probability of successful state flips.
A alternative methods to construct the attack pulses, involves retrieving publicly available calibration data from the hardware backend. While this approach reduces experimental overhead, it introduces a potential trade-off in precision, as the calibration data is typically updated once daily, and may not reflect real-time hardware conditions. Further, such approach can be easily defeated by the provider not publishing the needed calibration data; our approach makes no assumptions about calibration data that is public.

\subsection{Attack Deployment} \label{subsec:atk_deploy}
Following the attack pulse design for QubitHammer, the next step is deployment. QubitHammer can be deployed in three ways by the adversary, as described below.

\subsubsection{Single Attack Pulse Method}

For a single-pulse attack, the adversary crafts one microwave pulse to deploy on their drive channel using the victim’s calibrated resonance frequency ($f_0$) and drive amplitude ($A_0$). The pulse follows the same Gaussian (or DRAG-corrected) envelope from calibration and is truncated to a duration ($T_\theta$) that enacts the desired rotation angle $\theta$. By tuning $T_\theta$, the adversary can perform anything from a full $\pi$ rotation to a subtle partial flip, depending on the intended effect. 
Figure~\ref{subfig:single_pulse_wave} depicts a single attack pulse with a duration $T_\theta = 500dt$ \footnote{
$dt$ is the system cycle time that defines the frequency of quantum operations, and $dt = 2.22 ns$ is defined by the backend.}.

\subsubsection{Repeated Attack Pulse Method}

For a repeated-pulse attack, the adversary concatenates multiple calibrated Gaussian pulses, each tuned to the victim’s extracted frequency ($f_0$) and truncated to a desired duration $T_\theta$ on the adversary's drive line after inserting a precise delay $\delta t$ between successive attack pulses. This periodic sequence can either accumulate coherent rotations (multiplying the single-pulse angle $\theta$ by the number of pulses) or distribute small-angle drives to induce gradual decoherence and phase errors. A repeated pulse attack can enable richer adversarial strategies, from systematic bit-flip bursts to engineered leakage or subtle phase kicks. 
An example pulse-level representation of repeated attack pulses is depicted in Figure~\ref{subfig:multi_pulse_wave}, where the total duration of the repeated attack pulses is 2500$dt$. Each attack pulse has a duration of 500$dt$ which are separated by inserting uniform delays.

\subsubsection{Mixed Attack Pulse Method}

\begin{figure*}[t]

 \begin{subfigure}[b]{0.24\textwidth}
 \includegraphics[width=\textwidth, height = 2cm]{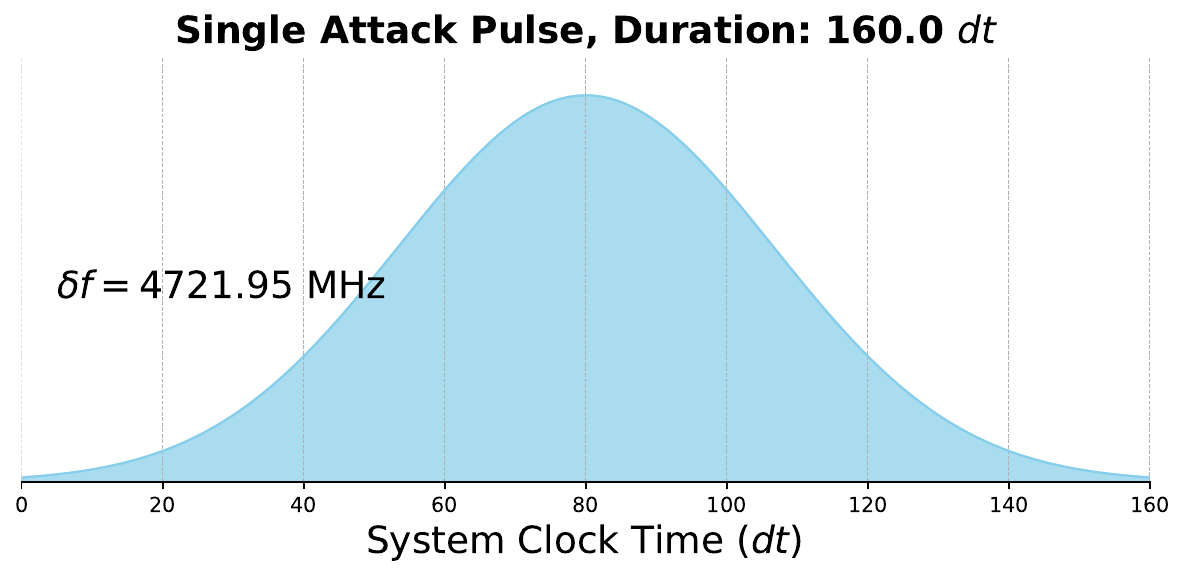}
 \caption{\small Single attack pulse.}
 \label{subfig:single_pulse_wave}
 \end{subfigure}
 \begin{subfigure}[b]{0.38\textwidth}
 \includegraphics[width=\textwidth, height = 2cm]{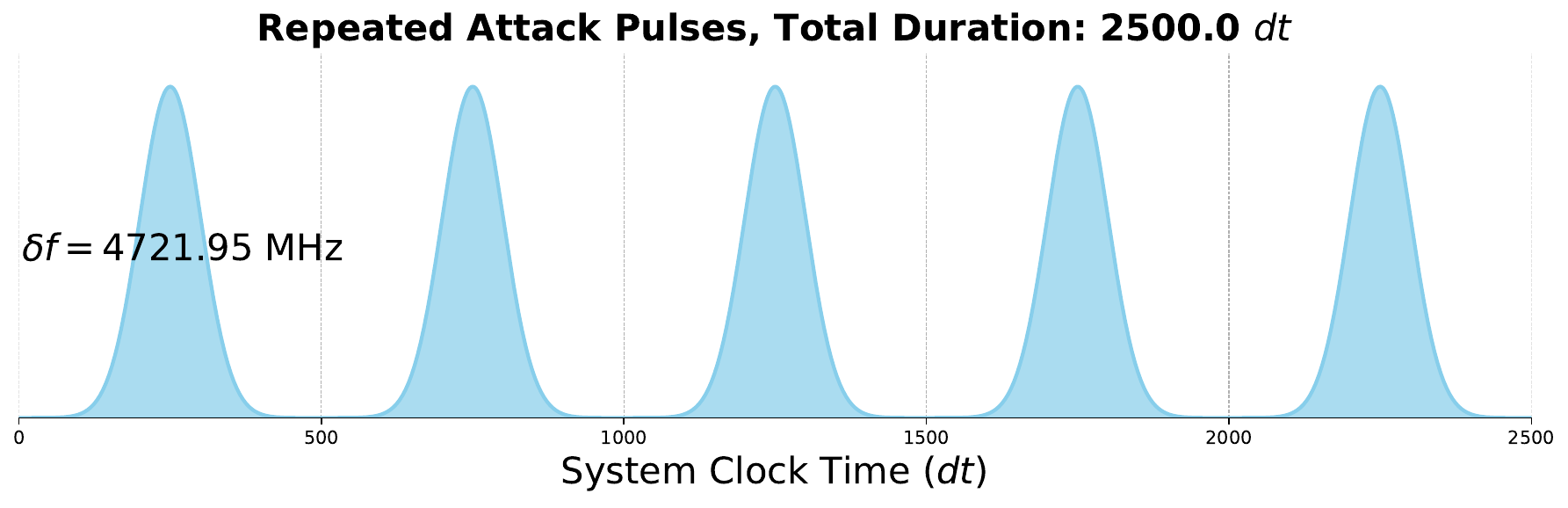}
 \caption{\small Repeated attack pulses.}
 \label{subfig:multi_pulse_wave}
 \end{subfigure}
 \begin{subfigure}[b]{0.36\textwidth}
 \includegraphics[width=\textwidth, height = 2cm]{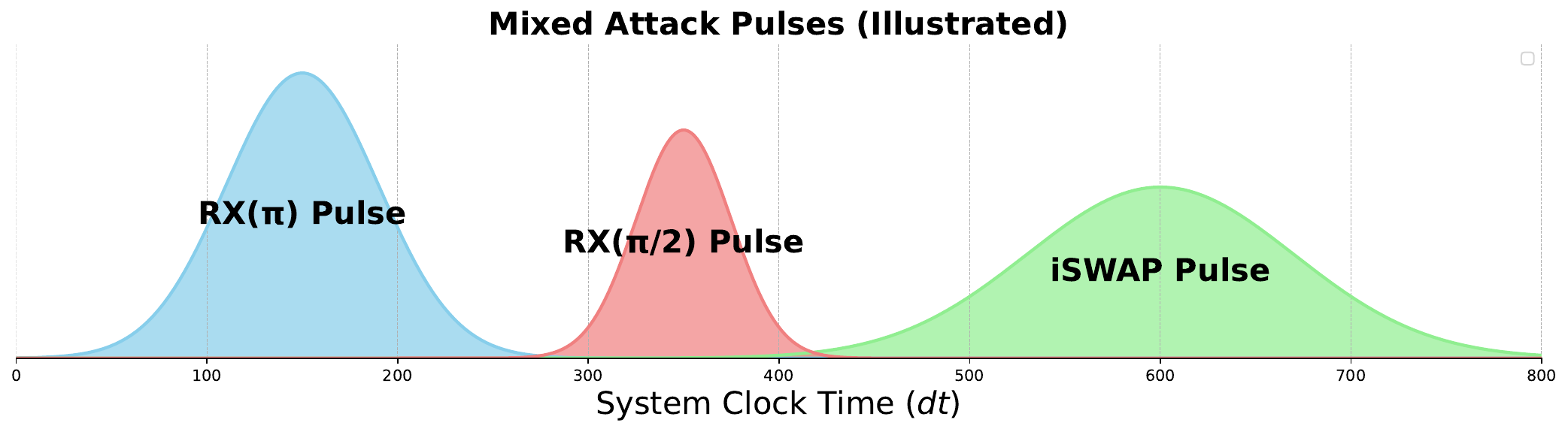}
 \caption{\small Mixed attack pulses.}
 \label{subfig:mixed_pulse_wave}
 \end{subfigure}
 \vspace{-1mm}
 \caption{\small An illustration of the different attack pulse visualizations. Please note the different cycle times on the x-axis as the figures are not drawn to scale. This is a conceptual diagram. The underlying physical implementation of these gates, particularly for the two-qubit {\tt iSWAP} operation, is more complex than shown here.}
 \label{fig:attack_pulse_waves}
\end{figure*}

For a mixed-pulse attack, the adversary constructs a set of calibrated native gate pulses \{($f_i$, $A_i$, $T_{\theta_i}$)\}. The adversary can select any pulses from the native gate set, implement them one by one and decide the sequences of these pulses himself.
Each pulse ($i$) in this sequence is calibrated to mimic a specific native gate operation, with its frequency ($f_i$), amplitude ($A_i$), and duration ($T_{\theta_i}$) precisely matching the parameters of the chosen gate.
This approach marks a significant advancement over the single- and repeated-pulse methods. It gives the adversary a wider range of attack pulse options. This flexibility is important because the effectiveness of any single pulse type, like the standard {\tt X}-gate, is not always reliable. The mixed-pulse method allows the adversary to choose the pulse that causes the most damage in a given situation, which increases the overall impact of the attack.
An example mixed attack pulses composed by three native gate pulses is illustrated in Figure~\ref{subfig:mixed_pulse_wave}. Start with a {\tt RX($\pi$)} pulse, then a {\tt RX($\pi$/2)} is applied, followed  by an {\tt iSWAP} pulse. Each is tuned according to their specific calibration data.

%% file: sections/attack_types.tex
\section{QubitHammer Scenarios} \label{sec:scenario}
\label{sec:attack}

\textcolor{black}{
QubitHammer targets multi-tenant superconducting processors where adversary and victim control disjoint qubit subsets on the same device. As in typical cloud settings, qubit assignments and physical placement are managed by the provider, influencing crosstalk and coupling pathways. To evaluate QubitHammer under varying conditions, we define four deployment scenarios differing in adversary qubit count and proximity to victim qubits. The specifics of each scenario are detailed subsequently.
}


\subsection{Attack Scenario 1: Short-Range Adversarial Impact with Proximal Qubit Allocation}

\textcolor{black}{
This scenario models a threat where the victim and adversary are allocated physically proximate qubits, potentially sharing a direct coupler. However, their programs remain logically isolated, with no shared gates or direct operational overlap. We evaluate two distinct cases, which differ based on the resources allocated to the adversary.
}
\subsubsection{Bulk-Access Interference (BAI)}
In this scenario, the adversary is allocated a majority of the processor’s qubits by the cloud scheduler, while the victim controls only a few. 
An extreme example is shown in Figure~\ref{fig:attack_type1}.
With extensive qubit access, the adversary can utilize attack pulses on one or more of its registers (highlighted in red). This attack scenario captures the impact of QubitHammer when the adversary has significantly favorable conditions, with a large number of qubits to deploy attack pulses, in addition to being co-located to the victim's qubits in the logical topology of the quantum computer, increasing chances of interference using control lines. 
The effectiveness of this attack scenario, including its potential to consistently disrupt quantum computations, is presented in Section~\ref{subsec:evaluation_1}.



\subsubsection{Constrained-Access Interference (CAI)}

The adversary is allocated a limited number of qubits that are still positioned close to the victim's qubits on the logical qubit topology as presented in Figure~\ref{fig:attack_type2}.
Although the adversary controls fewer qubits in this situation, the physical proximity of these qubits to the victim qubit on the quantum processor increases the likelihood of a successful attack. 
Even with a reduced number of qubits, the adversary can strategically apply pulses that exploit these interactions, potentially causing significant disruptions to the victim qubit’s state. The impact and effectiveness of this attack configuration are thoroughly examined in Section~\ref{subsec:evaluation_2}.

\subsection{Attack Scenario 2: Long-Range Adversarial Impact with Distant Qubit Allocation} \label{subsec:atk_3}
\textcolor{black}{
This scenario investigates the limits of spatial separation as a defense. The victim and adversary are separated by a significant physical buffer (at least eight qubits on thelogical topology). The purpose is to test QubitHammer's efficacy under conditions that are unfavorable to the adversary, specifically when direct proximity cannot be exploited. As with the previous scenario, we evaluate two distinct cases.
}
\subsubsection{Bulk-Access Interference (BAI)}

We consider a physically distant yet well-resourced adversary. Here, the victim qubits are separated from the adversary's by a significant buffer of at least eight logical qubits. However, the adversary compensates for this distance by controlling a large portion of the processor, potentially up to half its qubits, as illustrated in the extreme case in Figure~\ref{fig:attack_type3}.
The evaluation of this scenario is important because, despite the topological separation between the victim and adversarial qubits, the adversary's control over a large portion of the processor might still pose a significant threat. The adversary can deploy attack pulses across their allocated qubits in a coordinated manner, potentially creating long-range interference effects. These effects, while less direct due to the physical distance, can still induce errors in the victim's qubits, particularly through mechanisms such as resonant frequency overlap, crosstalk across the processor, or even indirect interactions through the quantum processor's control systems.
It is worth noting that this scenario raises important considerations about the effectiveness of attack strategies when the qubits under adversarial control are physically distant from the victim qubits in the quantum hardware. The extent to which long-distance perturbations can influence the victim’s operations, and how these interactions can be exploited to degrade performance, are key points of analysis that are explored in Section~\ref{subsec:evaluation_3}. 

\subsubsection{Constrained-Access Interference (CAI)}

In the final scenario, we consider both the victim and adversary to have access to a relatively small number of qubits (e.g., $10$ to $20$) as shown in Figure~\ref{fig:attack_type4}, which are physically separated by a significant distance in the quantum processor.
Although the significant physical separation between the qubits reduces the likelihood of direct interference, and the adversary has access to a limited number of qubits, the adversary can still potentially exploit long-range interactions, such as resonant frequency overlap or indirect coupling, to affect the victim's qubits. 
This scenario illustrates the possibility that even when the adversary controls only a modest number of qubits and maintains substantial logical separation from the victim’s registers, QubitHammer can nonetheless be executed successfully to compromise the victim’s computation.
The effectiveness of this long-distance attack and its potential to disrupt the victim’s computation is presented in detail in Section~\ref{subsec:evaluation_4}. 

\begin{figure*}[t]
 \centering
 \begin{subfigure}{0.25\linewidth}
 \centering
 \includegraphics[height = 3cm]{figures/attack_type1-compressed.pdf}
    \caption{\small Attack Scenario 1: \textit{BAI}}
    \label{fig:attack_type1}
 \end{subfigure}
 \begin{subfigure}{0.25\linewidth}
 \centering
 \includegraphics[height = 3cm]{figures/attack_type2-compressed.pdf}
    \caption{\small Attack Scenario 1: \textit{CAI}}
    \label{fig:attack_type2}
 \end{subfigure}
 \begin{subfigure}{0.24\linewidth}
 \centering
 \includegraphics[height = 3cm]{figures/attack_type3-compressed.pdf}
    \caption{\small Attack Scenario 2: \textit{BAI}}
    \label{fig:attack_type3}
 \end{subfigure}
 \begin{subfigure}{0.24\linewidth}
 \centering
 \includegraphics[height = 3cm]{figures/attack_type4-compressed.pdf}
    \caption{\small Attack Scenario 2: \textit{CAI}}
    \label{fig:attack_type4}
 \end{subfigure}
 \vspace{-5mm}
 \caption{\small Four attack scenarios introduced in this work. Both the adversary and the victim are allocated distinct and non-overlapping sets of qubits. The figures illustrate examples of attack scenarios on the IBM QPU with a single victim (green), and adversary (red). 
 }
 \label{fig:attack_types}
\end{figure*}

%% file: sections/exp_setup.tex
\section{Experimental Setup}
\label{sec:set_up}

This section details the experimental setup used to evaluate QubitHammer. Our experiments consider a multi-tenant environment where one victim and one adversary are co-located on a shared quantum processor. Consistent with attack scenatios in Section~\ref{sec:scenario}, both the victim and adversary can be allocated multiple, non-overlapping qubits, termed the "victim qubits" and "adversarial qubits", respectively.


\subsection{Hardware Utilized}

To evaluate QubitHammer under realistic scenarios defined in Section~\ref{sec:scenario} and within the real-world threat constraints of Section~\ref{sec:threat}, we conducted experiments on two state-of-the-art superconducting platforms. First, we utilize three of IBM Quantum’s 127-qubit Eagle r3 backends available for public access, namely, \textit{ibm\_brisbane}, \textit{ibm\_osaka} (retired August 13, 2024), and \textit{ibm\_kyoto}, and present the evaluation results in Section~\ref{sec:ibm_res}. Second, we validated QubitHammer on Rigetti’s 82-qubit Ankaa-class processor (\textit{Ankaa-3}) and detailed in Section~\ref{sec:rigetti_res}, to determine the cross-vendor applicability and intensity. It is important to emphasize that, although both IBM Quantum and Rigetti platforms are built on fixed-frequency transmon qubits, they differ substantially in coupling topology, microwave control electronics, packaging, and cryogenic wiring. By evaluating QubitHammer on varying architectures,  we aim to demonstrate that it can induce significant error rates across both heavy-hexagon and square-lattice superconducting qubit systems, underscoring the attack’s severity and hardware-agnostic potency.
\subsection{Attack Pulses}
In this section, we describe the attack pulses utilized to deploy QubitHammer on real-world quantum computers (IBM and Rigetti), following Section~\ref{subsec:pulse_design}. 
The attack pulses can be deployed on the adversarial qubits in two ways, as discussed in Section~\ref{subsec:atk_deploy}.

\subsubsection{Single and Repeated Attack Pulses on IBM}

Following Section~\ref{subsec:pulse_design}, for IBM hardware, we design the attack pulses using the frequency and amplitude of the {\tt X} gate for different victim qubits.
We set the duration of the attack pulse to 160 $dt$, which is the default value for IBM quantum systems.
It is important to note that the adversary can vary the number of attack pulses driving the adversarial qubits, as well as the duration of each attack pulse. 

\subsubsection{Single, Repeated, and Mixed Attack Pulses on Rigetti}
\label{subsec:rigetti_pulse}
While the Rigetti platform does not permit direct frequency sweeps like IBM, the users can programmatically access the daily calibration data for the victim qubit's native gates {\tt RX}, {\tt RZ}, and {\tt iSWAP} (Section~\ref{subsec:pulse_design}). Thus, the adversary can synthesize attack pulses by directly using the victim's precise parameters (frequency, amplitude, duration). This capability not only allows us to extend our evaluation from {\tt X}-gate pulse to those based on {\tt RX($\pi$/2)} and {\tt iSWAP}, but also facilitates the mixed pulse attack, where features from these different gate pulses are combined into a single adversarial circuit.


\subsection{Evaluation Metrics}

We quantify QubitHammer’s effect by measuring the variational distance between the baseline (attack-free) output distribution and the one produced under attack. Also known as the total variation distance $D_{TV}$, this metric is ideal here because it provides a comprehensive, norm‐based measure that captures the total statistical deviation between two output distributions induced by QubitHammer. It is defined as:
\vspace{-1.5mm}
\begin{equation} \label{eq:var_distance}
D_{\text{TV}}(P_{ideal}, P_{attacked}) = \frac{1}{2} \sum_{x \in X} \left| P_{ideal}(x) - P_{attacked}(x) \right|
\end{equation}

In Equation~\ref{eq:var_distance}, \(P_{ideal}(x)\) and \(P_{attacked}(x)\) represent the probabilities of the outcomes \(x\) under the ideal distributions with no attack and the distribution following the QubitHammer attack, respectively. The summation is taken over all possible outcomes \(x\) in the sample space \(X\). The factor of \(\frac{1}{2}\) ensures that the variational distance ranges from 0 to 1.
As an evaluation metric, the variational distance quantifies the maximum discrepancy between two possibility distributions: the victim's baseline output and its output in the presence of the QubitHammer attack. A distance ranging from 0 to 0.2 indicates a minimal impact of the attack, while a distance between 0.2 and 0.4 suggests a mild influence. A distance of 0.4 to 0.6 implies a significant impact, and a distance from 0.6 to 1 denotes a very high impact of the attack.
To provide a practical interpretation, Grover’s algorithm is applied to accelerate the identification of lead compounds by efficiently searching large chemical libraries for candidate molecules with high binding affinity. An adversarial attack with a high impact that compromises the outcome could misdirect months of experimental effort, compromise patient safety, and jeopardize the integrity of clinical trials.

\subsection{Evaluation Benchmarks}

To evaluate QubitHammer, we employ three distinct benchmarks, showcasing the experimental impact for single-pulse attacks and the real-world impact of repeated- and mixed-pulse~attacks.

\textbf{Idle qubits.} This provides a clean baseline with all \textit{`0'} output to measure the fundamental physical impact of QubitHammer without interference from a running algorithm, primarily used for evaluating single-pulse attacks.

\textbf{Grover's algorithm.} As a critical benchmarking primitive on NISQ hardware, it exemplifies quantum amplitude amplification by combining entangling operations, phase inversion, and inversion-about-the-mean. It is the primary target for our repeated-pulse attacks.

\textbf{Bell state preparation circuit.} As a foundational routine for generating entangled states, its use of multiple native gate types makes it an ideal target for evaluating the mixed-pulse~attack.
\




%% file: sections/ibm_exp_results.tex
\section{QubitHammer Evaluation on IBM}
\label{sec:ibm_res}

In this section, we present extensive experimental results for QubitHammer, evaluating each attack scenario from Section~\ref{sec:scenario} individually on IBM Quantum hardware. We analyze two deployment modes from Section~\ref{subsec:atk_deploy}:the adversary using a \textit{single attack pulse} and the adversary using \textit{repeated attack pulses} on their qubits. 

\subsection{Evaluation of Attack Scenario 1 (BAI)}
\label{subsec:evaluation_1}

In this section, we evaluate QubitHammer with Attack Scenario 1 BAI, where the adversary has access to qubits close to the victim in the quantum processor. 

\subsubsection{Single Attack Pulse Method}

\begin{figure}[t]
    \centering
    \includegraphics[width=0.45\textwidth]{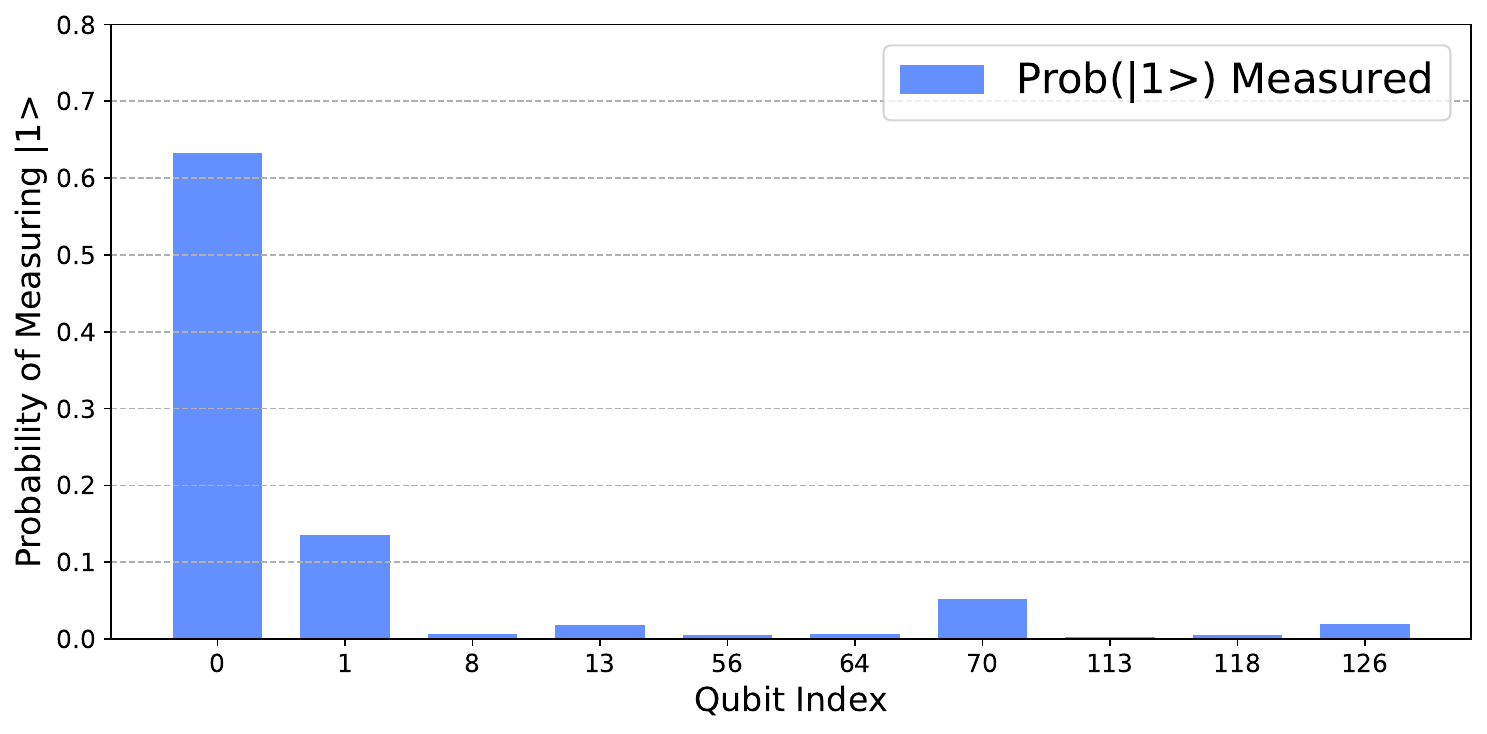}
    \vspace{-2mm}
    \caption{\small Qubit flipping probability from IBM \textit{\_brisbane} quantum processor with Attack Scenario 1 (CAI) using single pulse. Qubit index indicates the tested victim qubit. Only a limited number of victim qubits could be tested due to limited time available on this machine.}
    \label{fig:result_ibm}
\end{figure}

In this case, a single qubit is designated as the victim qubit, and all other qubits on the quantum processor are allocated to the adversary. 
Since each qubit has distinct properties such as coherence times and error rates, we first verify any dependence of attack potency on the victim qubit.
To this end, we examine the attack efficacy over 10 distinct qubits, situated across the topology of the IBM Eagle quantum processor, namely qubits 0, 1, 8, 13, 56, 64, 70, 113, 118, and 126. 
The results of our experiments for each victim qubit are highlighted in Figure~\ref{fig:result_ibm}. In these figures, the possible output states are denoted in the x-axis, and their corresponding obtained counts are shown in the y-axis. From the figure, it can be observed that there is a pronounced effect of the attack on victim qubit 0, shown in Figure~\ref{fig:result_ibm}. 
The variation distance was evaluated and summarized in Table~\ref{tab:attack_1_single_impact}. 
From the table, it can be observed that the attack is highly impactful on qubit 0, with a variational distance of 0.609, but shows low impact on the other qubits. However, upon more extensive evaluation, the same attack on Rigetti reveals additional qubit vulnerabilities, as discussed later in Section~\ref{sec:rigetti_res}.

\begin{figure}[t]
 \centering
 \begin{subfigure}[b]{0.45\linewidth}
 \includegraphics[width=\textwidth]{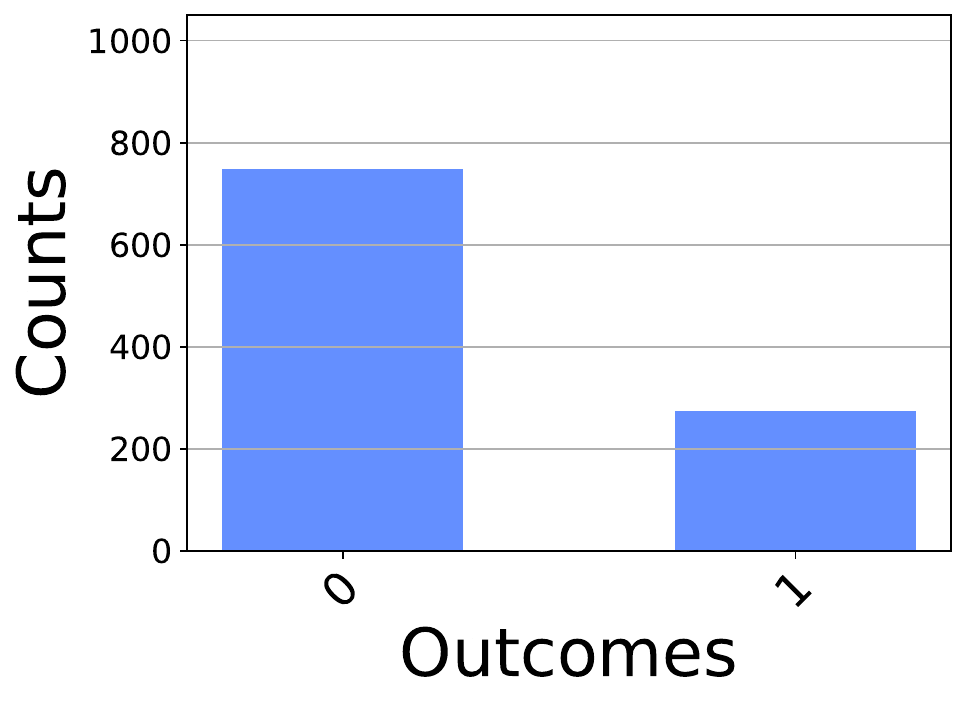}
 \caption{\small IBM\_kyoto}
 \label{subfig:kyoto_vic0_126att_single}
 \end{subfigure}
 \begin{subfigure}[b]{0.45\linewidth}
 \includegraphics[width=\textwidth]{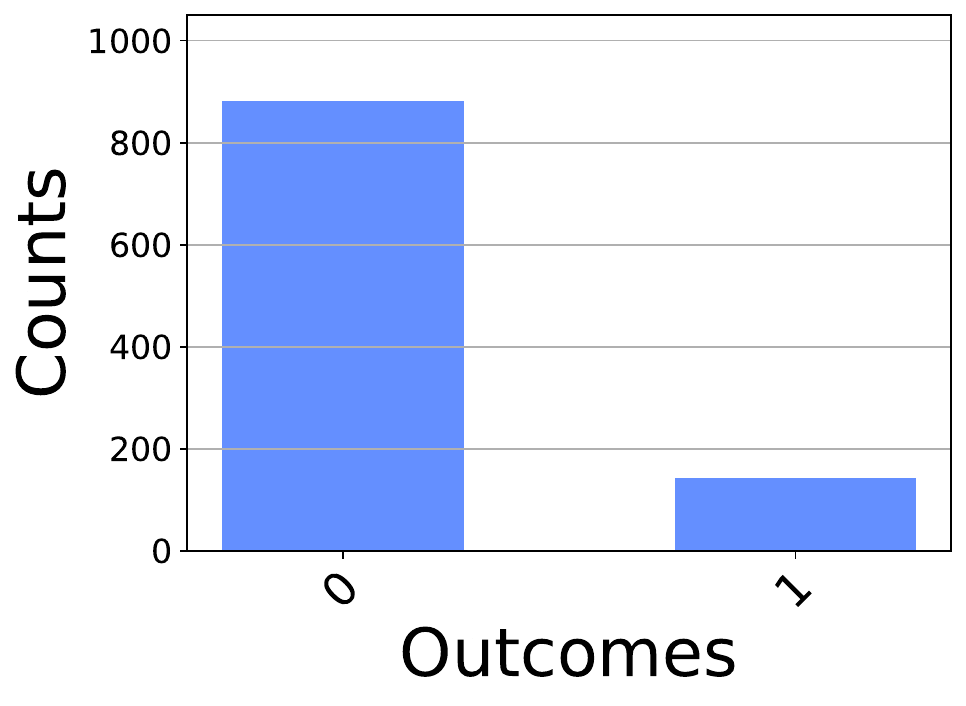}
 \caption{\small IBM\_osaka}
 \label{subfig:osaka_vic0_126att_single}
 \end{subfigure}
 \vspace{-1mm}
 \caption{\small Demonstration of Attack Scenario 1 (BAI) on additional IBM Quantum machines: \textit{IBM\_kyoto} and \textit{IBM\_osaka} on victim~qubit. Single pulse attack method is tested, targeting victim qubit 0.}
 \label{fig:result_kyoto_osaka_vic0_126att_single}
\end{figure}

To verify these findings across different hardware, we replicate the attack on \textit{IBM\_kyoto} and \textit{IBM\_osaka}, again using qubit 0 as the victim. 
The results, displayed in Figure~\ref{fig:result_kyoto_osaka_vic0_126att_single}, yielded variational distances of \textit{0.26} and \textit{0.131}, respectively, indicating minimal to moderate efficacy.
Similar to \textit{IBM\_brisbane}, the attack was much less effective on other victim~qubits.

\begin{table}[t]
\centering
\caption{\small Impact of Attack Scenario 1 (BAI) with a Single Pulse method, evaluated on \textit{IBM\_Brisbane} machine.}
\vspace{-1mm}
\renewcommand{\arraystretch}{2} 
\setlength{\tabcolsep}{3pt} 
\resizebox{\linewidth}{!}{%
\begin{tabular}{c|c|c|c|c|c|c|c|c|c|c}
\hline
\textbf{Victim qubit}         & 0     & 1     & 8     & 13    & 56    & 64    & 70    & 113    & 118    & 126    \\ \hline
\textbf{Variational Distance} & 0.609 & 0.123 & 0.003 & 0.009 & 0.004 & 0.003 & 0.042 & 0.0059 & 0.0029 & 0.0039 \\ \hline
\end{tabular}%
}
\label{tab:attack_1_single_impact}
\end{table}

\subsubsection{Repeated Attack Pulses Method}
\label{subsubsec:repeated_pulse}

\begin{figure}[t]
\centering
\includegraphics[width=0.45\textwidth]{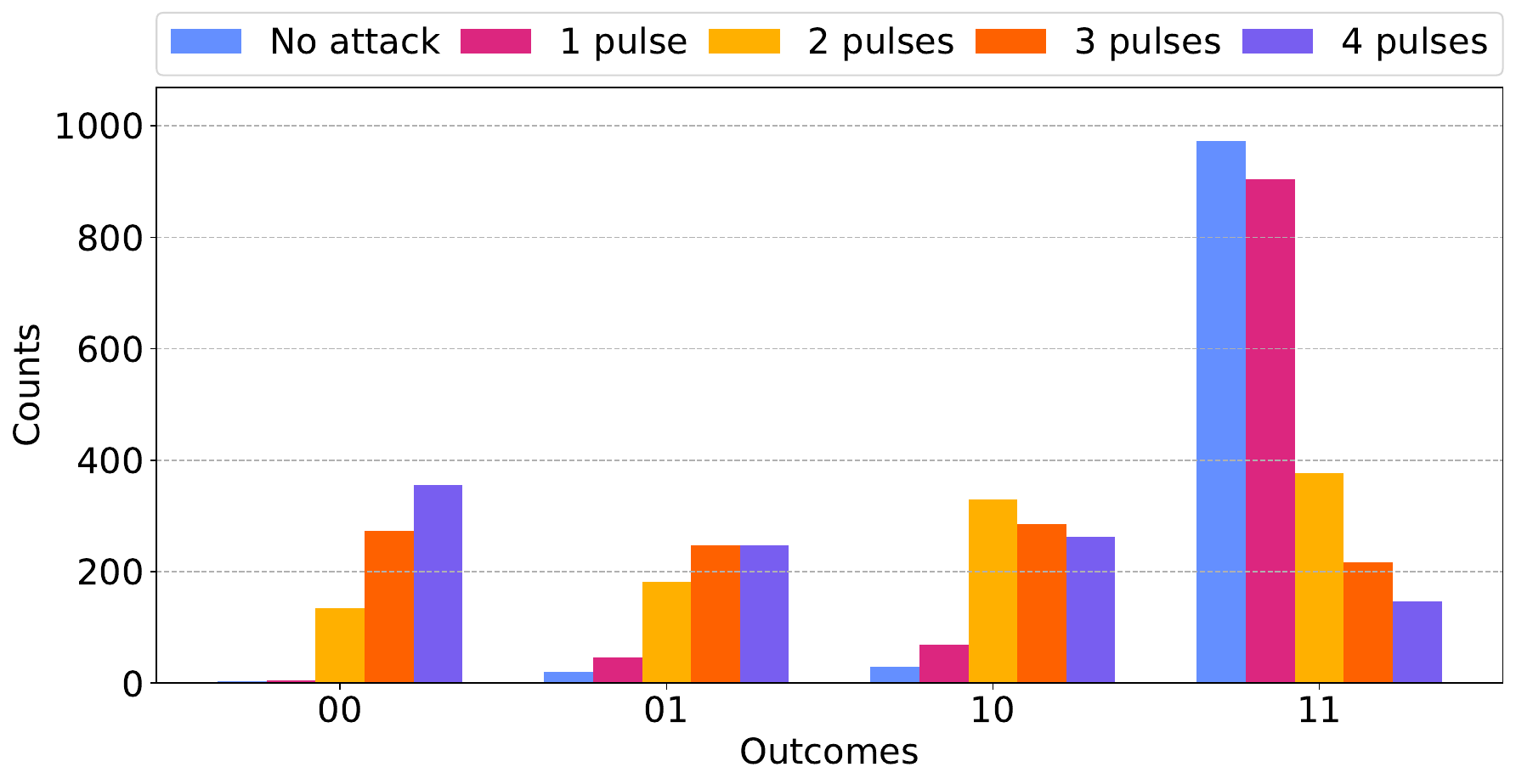}
\vspace{-2mm}
 \caption{\small Effect of varying the number of attack pulses used to drive adversarial qubits, on attack potency on Grover algorithms with Attack Scenario 1 (BAI) using qubit 0,1 on \textit{ibm\_brisbane}.}
 \label{fig:result_ibm_repeated}
\end{figure}

For this experiment, we allocate two qubits, 0 and 1, to the victim, and all other qubits on the quantum processor are assigned to the adversary. To accentuate the impact of QubitHammer in real-world scenarios, we consider the victim is running a two-qubit Grover's circuit on their qubits. The number of attack pulses on each adversarial qubit is gradually increased from one to four, and their impact on the output fidelity of the victim circuit is displayed in Figure~\ref{fig:result_ibm_repeated}. 
Upon evaluation, variational distances of \textit{0.074}, \textit{0.586}, \textit{0.741}, and \textit{0.801} were furnished for \textit{one}, \textit{two}, \textit{three}, and \textit{four} attack pulses, respectively. These results indicate that the attack potency of QubitHammer increases progressively with an increasing number of attack pulses employed by the adversary in this attack scenario, becoming highly effective using just four attack pulses.


\subsection{Evaluation of Attack Scenario 1 (CAI)}
\label{subsec:evaluation_2}
In this section, we evaluate QubitHammer with Attack Scenario 1 CAI, where the adversarial qubits are limited, but located in physical proximity to the victim's qubits on the quantum processor.

\subsubsection{Single Attack Pulse Method}

For this evaluation, victim qubit allocation remains the same as Section~\ref{subsec:evaluation_1} (\textit{i.e}., qubit 0 is allocated to the victim), while the number of adversarial qubits was varied from 8 to 67.
The results, depicted in Figure~\ref{subfig:result_vic0_dis1_8-67att}\footnote{The term "attacker" used in figure~\ref{subfig:result_vic0_dis1_8-67att}, ~\ref{fig:atk_3}, and ~\ref{subfig:result_vic0_dis9_18-32att}'s label corresponds to the "adversarial qubits" discussed in the text.} and summarized in Table~\ref{tab:attack_2_single_impact}, show a clear correlation between the number of adversaries and the attack's impact.
From the figure, it can be observed that as the number of adversaries increases, the output fidelity decreases noticeably.
From the table, it can be observed that the attack potency increases from minimal (variational distance of \textit{0.122} for 8 adversarial qubits) to highly effective (variational distance of \textit{0.75} for 60 adversarial qubits). These results highlight that a high number of adversarial qubits increases the efficacy of QubitHammer in this Attack Scenario.

\begin{table}[t]
\centering
\caption{\small Impact of Attack Scenario 1 (CAI) with a Single Pulse~method.}
\vspace{-1mm}
\renewcommand{\arraystretch}{2} 
\setlength{\tabcolsep}{4pt} 
\resizebox{\linewidth}{!}{%
\begin{tabular}{c|c|c|c|c|c|c|c|c}
\hline
\textbf{Number of adversaries}  & 8     & 10    & 15    & 19   & 23    & 30    & 60   & 67    \\ \hline
\textbf{Variational Distance} & 0.122 & 0.116 & 0.227 & 0.37 & 0.439 & 0.623 & 0.75 & 0.596 \\ \hline
\end{tabular}%
}
\label{tab:attack_2_single_impact}
\end{table}

\subsubsection{Repeated Attack Pulses Method}

\begin{figure}[t]
 \centering
 \begin{subfigure}[b]{0.49\linewidth}
 \includegraphics[width=\textwidth]{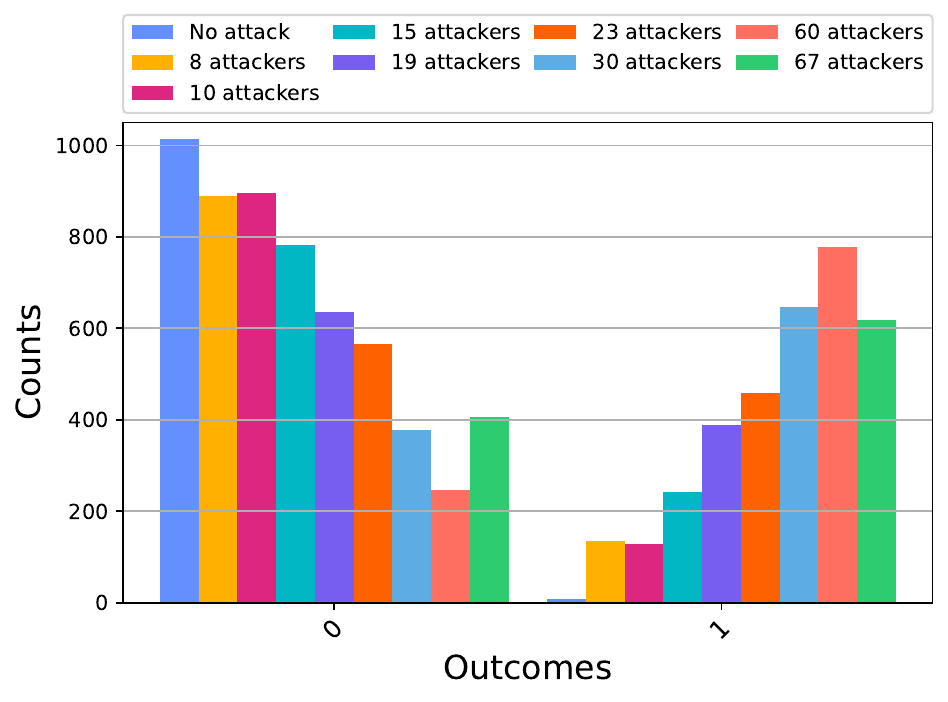}
 \caption{\small Single Pulse method.}
 \label{subfig:result_vic0_dis1_8-67att}
 \end{subfigure}
 \begin{subfigure}[b]{0.49\linewidth}
 \includegraphics[width=\textwidth]{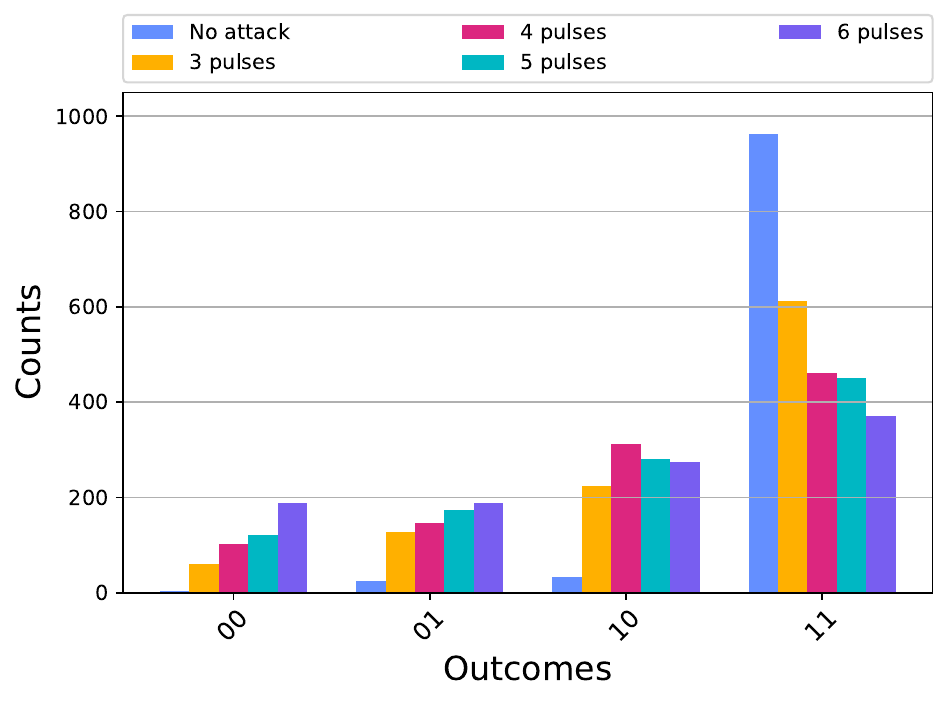}
 \caption{\small Repeated Pulses method.}
 \label{subfig:mid_attack}
 \end{subfigure}
 \vspace{-4mm}
\caption{\small Evaluation of Attack Scenario 1 (CAI) on \textit{IBM\_brisbane}.}
 \label{fig:sce1_cai}
\end{figure}

Similar to Section~\ref{subsubsec:repeated_pulse}, the victim is considered to run Grover's algorithm on qubits 0 and 1, while 32 qubits (5-36) are given to the adversary. The results in Figure~\ref{subfig:mid_attack} show the effect of increasing the number of attack pulses from three to six. 
Variational distances of \textit{0.342}, \textit{0.489}, \textit{0.501}, and \textit{0.578}, were obtained for 3, 4, 5 and 6 attack pulses, respectively. 
This demonstrates that even with a limited number of qubits under the adversary's control, QubitHammer can still significantly compromise the victim's outcome.


\subsection{Evaluation of Attack Scenario 2 (BAI)}
\label{subsec:evaluation_3}
Here, we evaluate the impact of Attack Scenario 2 with BAI, in which the victim and adversarial qubits are significantly separated on the quantum processor topology.


\subsubsection{Single Attack Pulse Method}

\begin{figure}[t]
 \begin{subfigure}[b]{0.49\linewidth}
 \includegraphics[width=\textwidth]{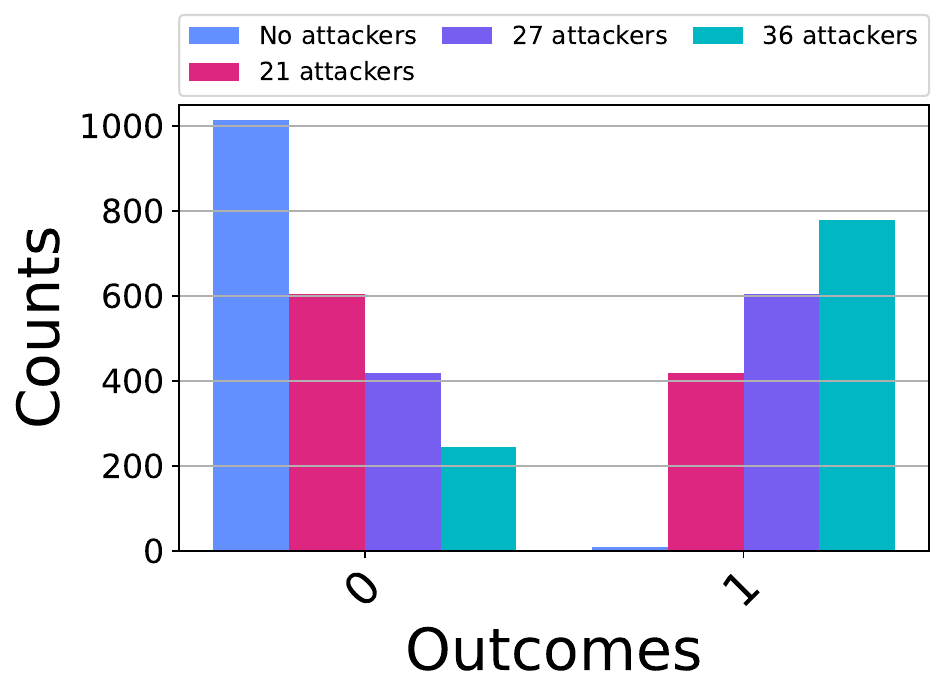}
 \caption{\small Adversarial qubitss at distance 13.}
 \label{subfig:result_vic0_dis13_21-36att}
 \end{subfigure}
  \begin{subfigure}[b]{0.49\linewidth}
 \includegraphics[width=\textwidth]{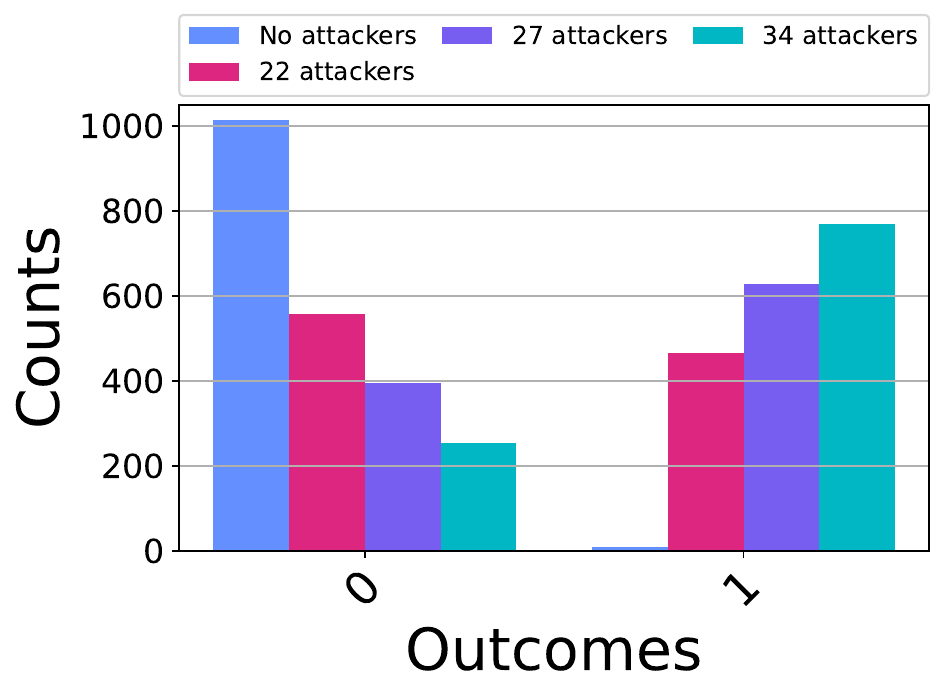}
 \caption{\small Adversarial qubitss at distance 17.}
 \label{subfig:result_vic0_dis17_22-34att}
 \end{subfigure}
 \caption{\small Evaluation results for Attack Scenario 2 (BAI) with varying number of adversarial qubits on \textit{IBM\_kyoto}.}
 \label{fig:atk_3}
\end{figure}

We keep the victim scenario consistent with Section~\ref{subsec:evaluation_1}, while varying the number of adversarial qubits. For this scenario, we fix the separation of the victim and adversary qubits at distances of 13 and 17, to assess the impact of the separation of victim and adversary qubits in the processor topology. The results presented in Figure~\ref{fig:atk_3} show that the attack's efficacy increases with the number of adversaries, at both distances of 13 and 17. 
Figure~\ref{subfig:result_vic0_dis13_21-36att} shows, at a distance of 13, increasing the adversarial qubits from 21, 27 to 36 raised the variational distance from \textit{0.401}, \textit{0.581} to \textit{0.753}.
Similarly, at a distance of 17, Figure~\ref{subfig:result_vic0_dis17_22-34att} depicts that using 22, 27, and 34 adversarial qubits yeilds increasing variational distances from \textit{0.445}, \textit{0.604} to \textit{0.743}. 
These results confirm that the effect of QubitHammer becomes more profound with increasing the number of qubits controlled by the adversary, independent of separation on the processor.

\subsubsection{Repeated Attack Pulses Method} \label{subsubsec:three_multi_pulse}

\begin{figure}[t]
 \centering
 \begin{subfigure}[b]{0.49\linewidth}
 \includegraphics[width=\textwidth]{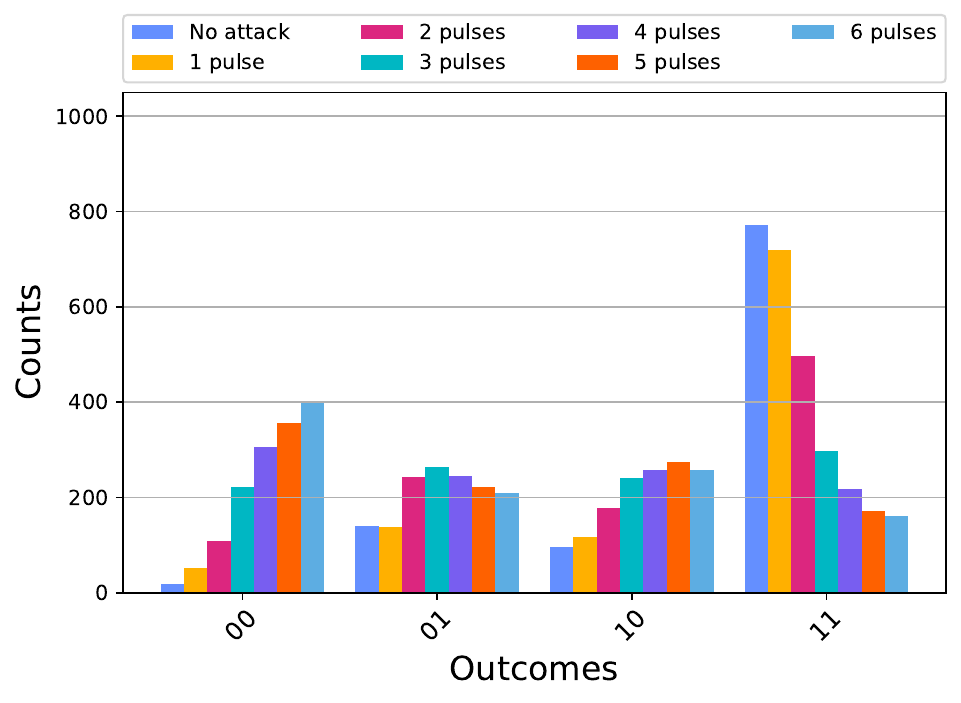}
 \caption{\small Evaluation on \textit{IBM\_kyoto}}
 \label{subfig:result_grover_kyoto_vic0_att71-126_pulse0-6}
 \end{subfigure}
 \begin{subfigure}[b]{0.49\linewidth}
 \includegraphics[width=\textwidth]{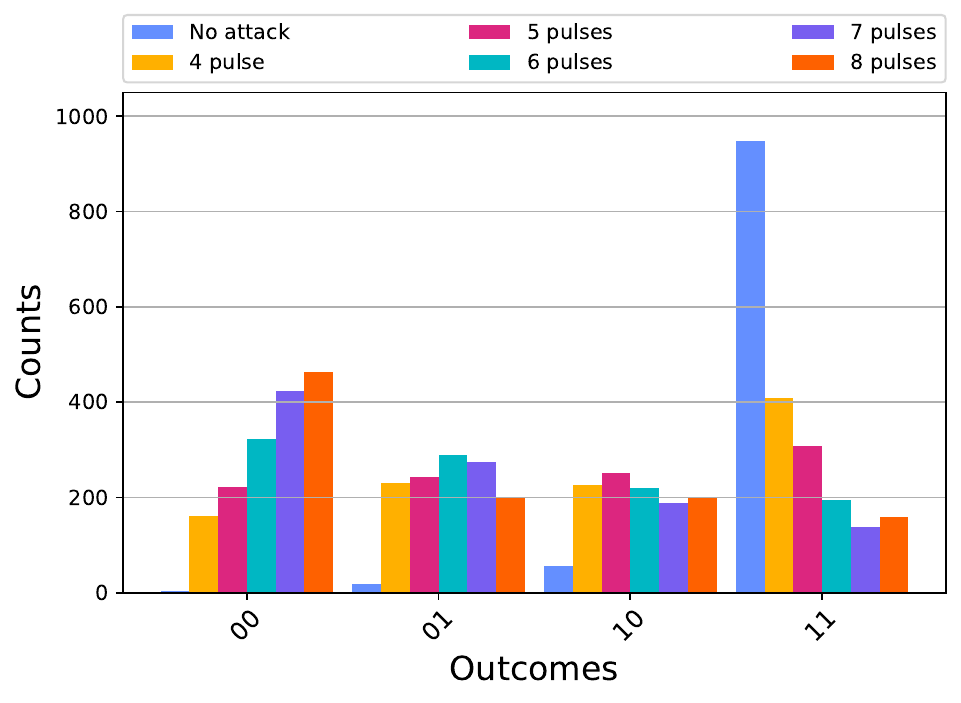}
 \caption{\small Evaluation on \textit{IBM\_osaka}.}
 \label{subfig:result_grover_osaka_vic0_att71-126_pulse04-8}
 \end{subfigure}
 \vspace{-4mm}
\caption{\small Evaluation of Attack Scenario 2 (BAI).}
 \label{fig:sce2_bai}
\end{figure}

We then test a repeated-pulse attack with victim setup similar to Section~\ref{subsubsec:repeated_pulse}  on \textit{IBM\_kyoto} and \textit{IBM\_osaka}. The adversary is assigned qubits 71-126, which are at a distance of 13 from the victim qubits on the processor.
As shown in Figures~\ref{subfig:result_grover_kyoto_vic0_att71-126_pulse0-6} and~\ref{subfig:result_grover_osaka_vic0_att71-126_pulse04-8}, increasing the number of attack pulses progressively degraded the circuit's output fidelity on both systems. 
For \textit{IBM\_kyoto}, variational distances of \textit{0.052, 0.269, 0.462, 0.541}, and \textit{0.597} were obtained for \textit{two} to \textit{six} attack pulses, in increasing order as shown in Figure~\ref{subfig:result_grover_kyoto_vic0_att71-126_pulse0-6}.
For \textit{IBM\_osaka}, Figure~\ref{subfig:result_grover_osaka_vic0_att71-126_pulse04-8} illustrates variational distances grow from \textit{0.526}, \textit{0.624}, \textit{0.734}, \textit{0.791} to \textit{0.770} as attack pulses increase from \textit{four} to \textit{eight}.

\subsection{Evaluation of Attack Scenario 2 (CAI)}
\label{subsec:evaluation_4}
In this section, we assess the final Attack Scenario, where the adversary controls a moderate number of qubits (fewer than 35) and is positioned at a moderate distance (15-20) from the victim qubits within the quantum processor. 

\subsubsection{Single Attack Pulse Method}

We assign qubit 0 to the victim and varied the number of adversarial qubits from 18 to 32 at a distance of 9 from the victim. The results, summarized in Figure~\ref{subfig:result_vic0_dis9_18-32att}, confirms that the attack's impact scales with the number of adversarial qubits. This is corroborated by the evaluation results, where variational distances of \textit{0.357}, \textit{0.508}, and \textit{0.688} were obtained for \textit{18}, \textit{23}, and \textit{32} adversarial qubits, respectively.
Notably, it can be observed that using 23 or more adversarial qubits causes the victim’s circuit to produce erroneous output.

\subsubsection{Repeated Attack Pulses Method}

\begin{figure}[t]
 \centering
 \begin{subfigure}[b]{0.49\linewidth}
 \includegraphics[width=\textwidth]{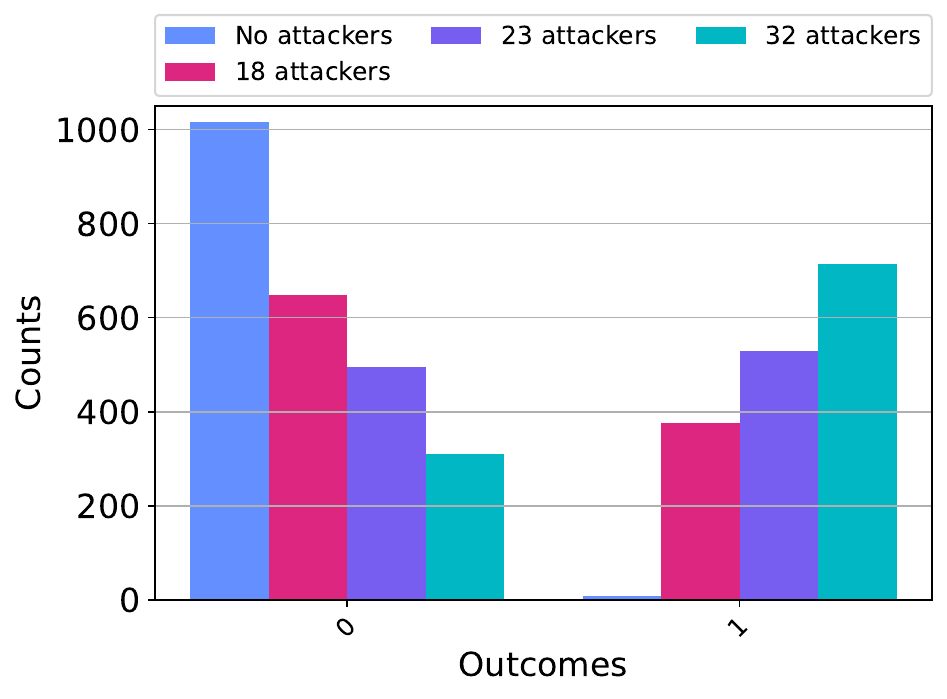}
 \caption{\small Single attack pulse.}
 \label{subfig:result_vic0_dis9_18-32att}
 \end{subfigure}
 \begin{subfigure}[b]{0.48\linewidth}
 \includegraphics[width=\textwidth]{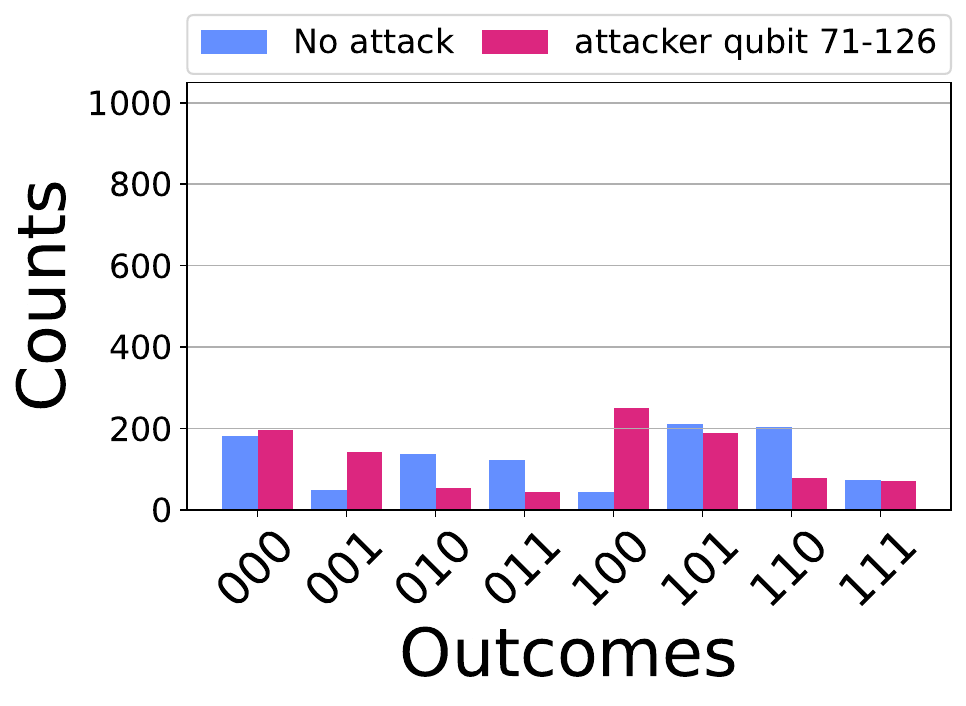}
 \caption{\small Repeated attack pulses.}
 \label{subfig:result_grover_vic01pair_71-126-86att}
 \end{subfigure}
\caption{\small Evaluation of Attack Scenario 2 (CAI) on \textit{IBM\_kyoto}.}
 \label{fig:sce2_cai}
\end{figure}

For this evaluation, we used a five-pulse attack against a Grover's algorithm on the victim's assigned qubits 0 and 1). We varied the number of adversarial qubits, starting from a large set (71-126) and dually reducing it to a smaller one (71-86).
The results in Figure~\ref{subfig:result_grover_vic01pair_71-126-86att} show that the attack's potency diminishes as the number of adversarial qubits is reduced.
Upon evaluation, variational distances of \textit{0.429, 0.501, 0.521, 0.600}, and \textit{0.664} were obtained for adversarial qubits in the range 71-86, 71-96, 71-106, 71-116, and 71-126, respectively. 
This indicates a significant impact of QubitHammer despite a limited number of adversarial qubits (for range 71-86, a variational distance of 0.429 was obtained).
This final scenario demonstrates that even when the adversary is constrained by both the number of qubits and the distance from the victim, QubitHammer still exerts considerable influence over the outcome of the victim's quantum operations, confirming a viable threat even with limited resources.

\subsection{Additional Evaluation}

\textcolor{black}{
We further test QubitHammer's robustness by applying it to highly optimized circuits and by targeting the QAOA algorithm. Evaluating against high optimization levels confirms that standard compiler defenses are insufficient against our targeted attack. Targeting QAOA provides a more comprehensive test than attacking Grover's algorithm. Its reliance on a full probability distribution allows us to demonstrate that QubitHammer can cause complex damage by distorting the entire outcome space, not merely causing bit-flips.
}

\subsubsection{QubitHammer on Highest Optimization}
\begin{figure}[t]
 \centering
 \begin{subfigure}[b]{0.45\linewidth}
 \includegraphics[width=\textwidth]{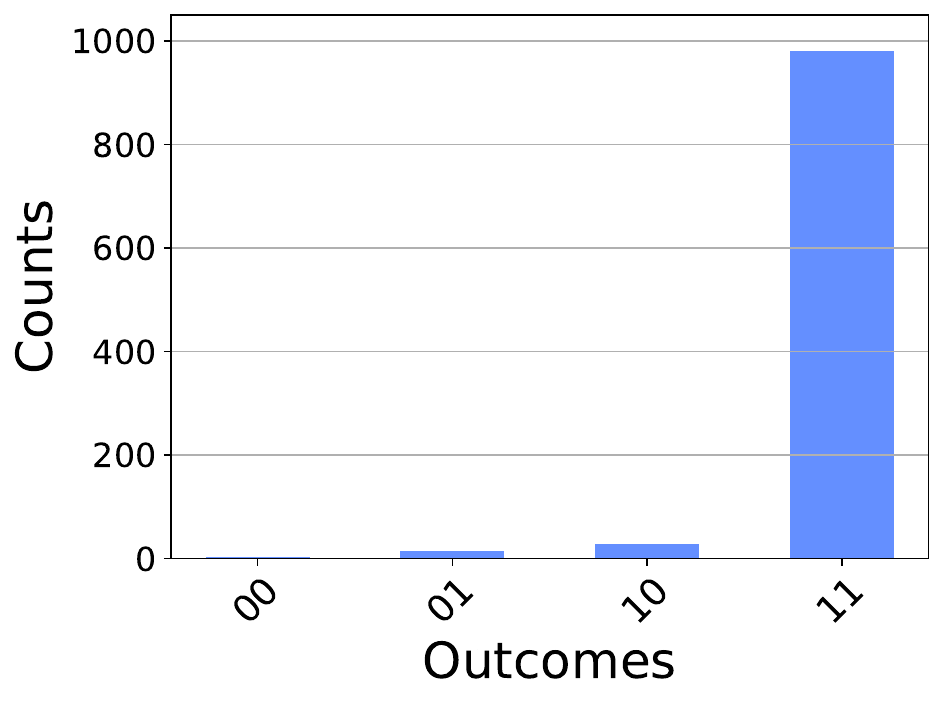}
 \caption{\small Grover circuit output with optimization level 3.}
 \label{subfig:res_5_1}
 \end{subfigure}
 \begin{subfigure}[b]{0.45\linewidth}
 \includegraphics[width=\textwidth]{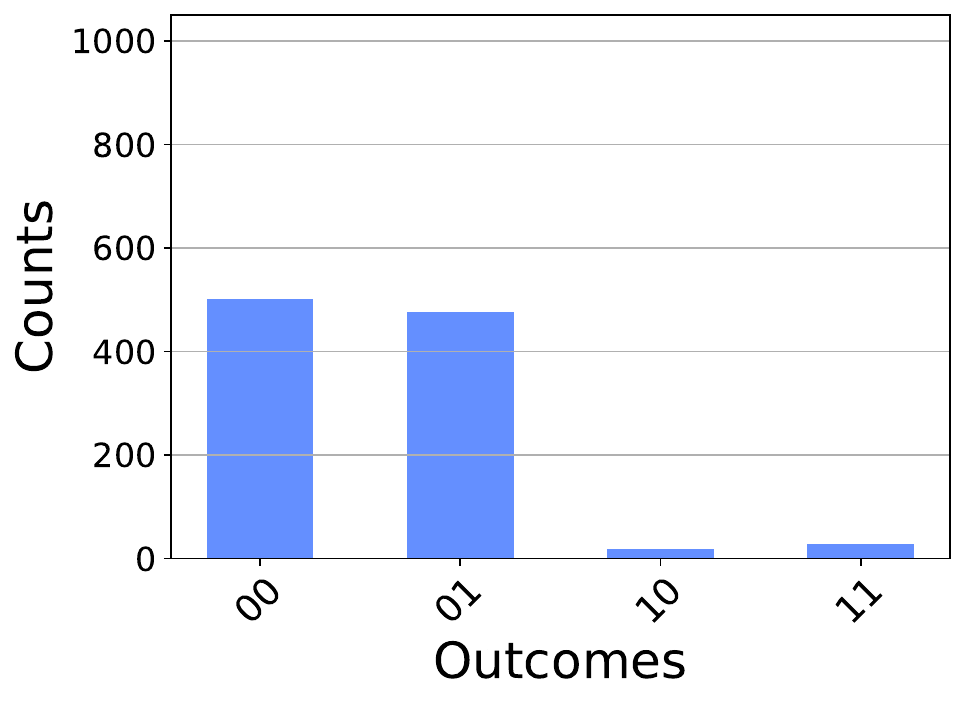}
 \caption{\small Output following attack on optimization level 3.}
 \label{subfig:res_5_2}
 \end{subfigure}
\caption{\small  Additional evaluation, efficacy of Attack Scenario 2 (BAI) on maximally optimized Grover's two-qubit circuit.}
 \label{fig:compiled_res_5}
\end{figure}

In the previous sections, all attacks were executed on victim circuits transpiled with optimization level 0 to demonstrate the efficacy of the attacks. 
To test the impact of QubitHammer on higher optimization levels during transpilation, we modify the victim's Grover circuit by using optimization level 3 offered by IBM's transpiler. In this experiment, qubits 0 and 1 were assigned to the victim, while qubits 71 to 126 were allocated to the adversary, with five attack pulses applied (attack scenario 3 (BAI) with repeated pulses).
The results of this experiment are presented in Figure~\ref{fig:compiled_res_5}, which depicts a completely skewed output distribution following the attack, with a variational distance of 0.938. This finding highlights that existing optimization techniques are rendered useless against QubitHammer.

\subsubsection{QubitHammer on QAOA}
\begin{figure}[t]
 \centering
 \begin{subfigure}[b]{0.48\linewidth}
 \includegraphics[width=\textwidth]{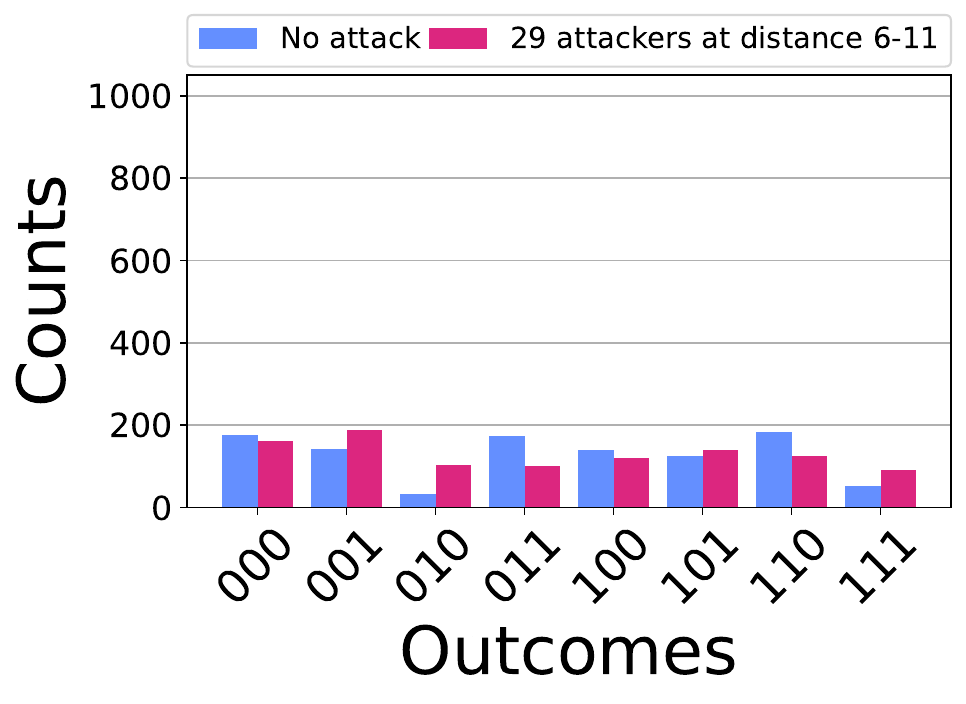}
 \caption{\small Attack Scenario 1 (CAI).}
 \label{subfig:res_qaoa_1}
 \end{subfigure}
 \begin{subfigure}[b]{0.48\linewidth}
 \includegraphics[width=\textwidth]{figures/experiment_result/result_qaoa_brisbane01_att71-126_4pulses.pdf}
 \caption{\small Attack Scenario 2 (BAI.}
 \label{subfig:res_qaoa_2}
 \end{subfigure}
 \vspace{-2mm}
\caption{\small Additional evaluation, efficacy of Attack Scenario 1 and 2 on a three-qubit QAOA circuit.}
 \label{fig:compiled_res_qaoa}
\end{figure}

In this section, we apply QubitHammer to a four-qubit QAOA algorithm being executed by the victim, on the \textit{IBM\_brisbane} quantum processor. For evaluation, we consider attack scenario 1 (CAI) and attack scenario 2 (BAI), with four attack pulses on each adversary qubit. The results, shown in Figures~\ref{subfig:res_qaoa_1} and~\ref{subfig:res_qaoa_2}, respectively, furnish variational distances of \textit{0.17} and \textit{0.307} indicating non-negligible impact of QubitHammer. Notably, even with only four adversarial pulses in Scenario 2 (BAI), QubitHammer achieved moderate efficacy, demonstrating its potency across different quantum algorithms despite resource constraints.

%% file: sections/rigetti_exp_results.tex
\section{QubitHammer Evaluation on Rigetti}
\label{sec:rigetti_res}
Our evaluation on Rigetti \textit{Ankaa-3} processor was enabled by its programmatic access to gate calibration data and extended access time, which allowed for two key extensions: the implementation of the mixed-pulse attack and a comprehensive full-chip evaluation.
The evaluation of this new attack method prompted us to organize this section by attack method rather than by deployment scenarios.

\subsection{Evaluation of Single Attack Pulse}

\begin{figure}[t]
    \centering
    \includegraphics[width=0.45\textwidth]{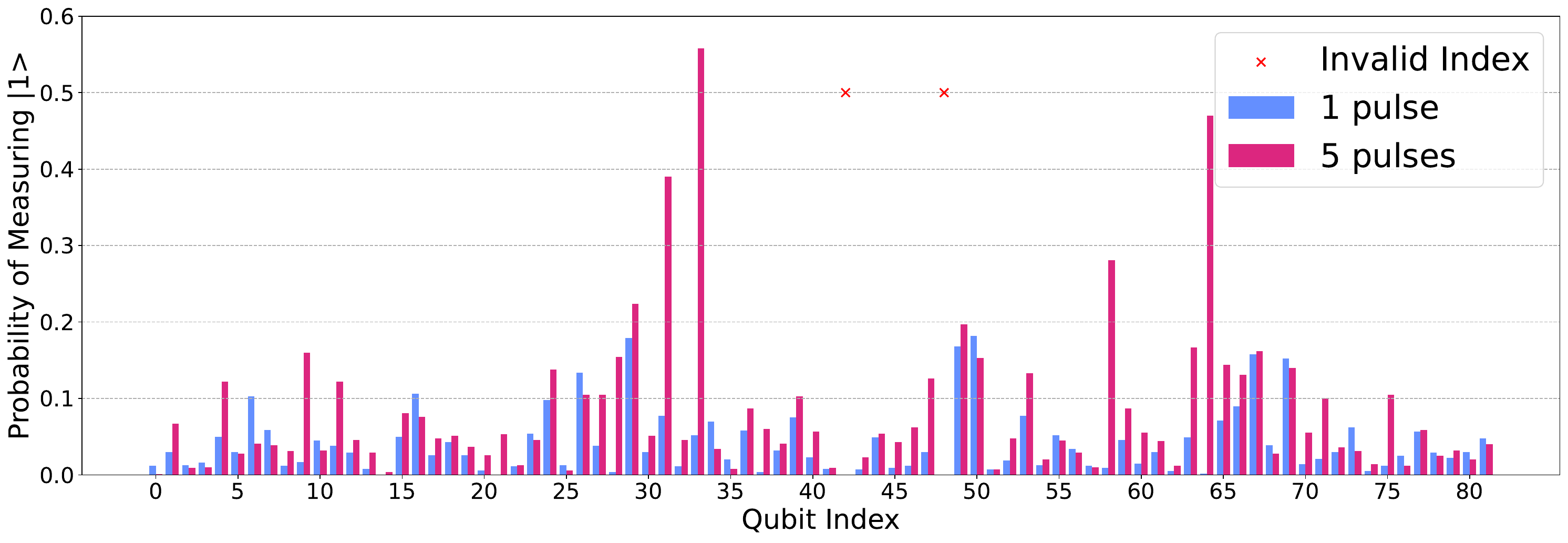}
    \vspace{-2mm}
    \caption{\small Qubit flipping probability from Rigetti \textit{Ankaa-3} quantum processor with Attack Scenario 1 (BAI) using single pulse and 5 pulses. All available qubits were tested.}
    \label{fig:result_rigetti_all_qubits}
\end{figure}

We conducted a full-chip evaluation on the Rigetti \textit{Ankaa-3}, with Attack Scenario 1 (BAI), where each idle qubit, in turn, served as the victim while the remaining 81 qubits . A single attack {\tt X}-gate pulse is performed on each of the adversarial qubits.
The result flipping probabilities, depicted in Figure~\ref{fig:result_rigetti_all_qubits}, vary significantly, with certain qubits (e.g., 29, 49, 50) showing heightened vulnerability to this single-pulse attack.
Notably, while the peak flipping probability on Rigetti (under 0.2) is considerably lower than that observed on the IBM system (over 0.6), the impact is more widespread, with a larger number of qubits affected by QubitHammer attack.

\subsection{Evaluation of Repeated Attack Pulse}

First, we run a full-chip evaluation of the repeated pulse attack: Each single idle qubit is allocated, in turn, to the victim, while the adversary apply 5 repeated {\tt X}-gate pulses on the remaining 81 adversarial qubits. As shown in Figure~\ref{fig:result_rigetti_all_qubits}, this repeated attack led to a significant increase in flipping probabilities across the entire chip compared to the single-pulse case. Although susceptibility still varied between qubits, the overall impact was much more severe. For example, the flipping probability for qubit 33 surged to 0.558, while other qubits such as 64, 58, and 31 also exhibited substantial error rates of 0.470, 0.390, and 0.281, respectively.

\begin{figure}[t]
\centering
\includegraphics[width=0.8\linewidth]{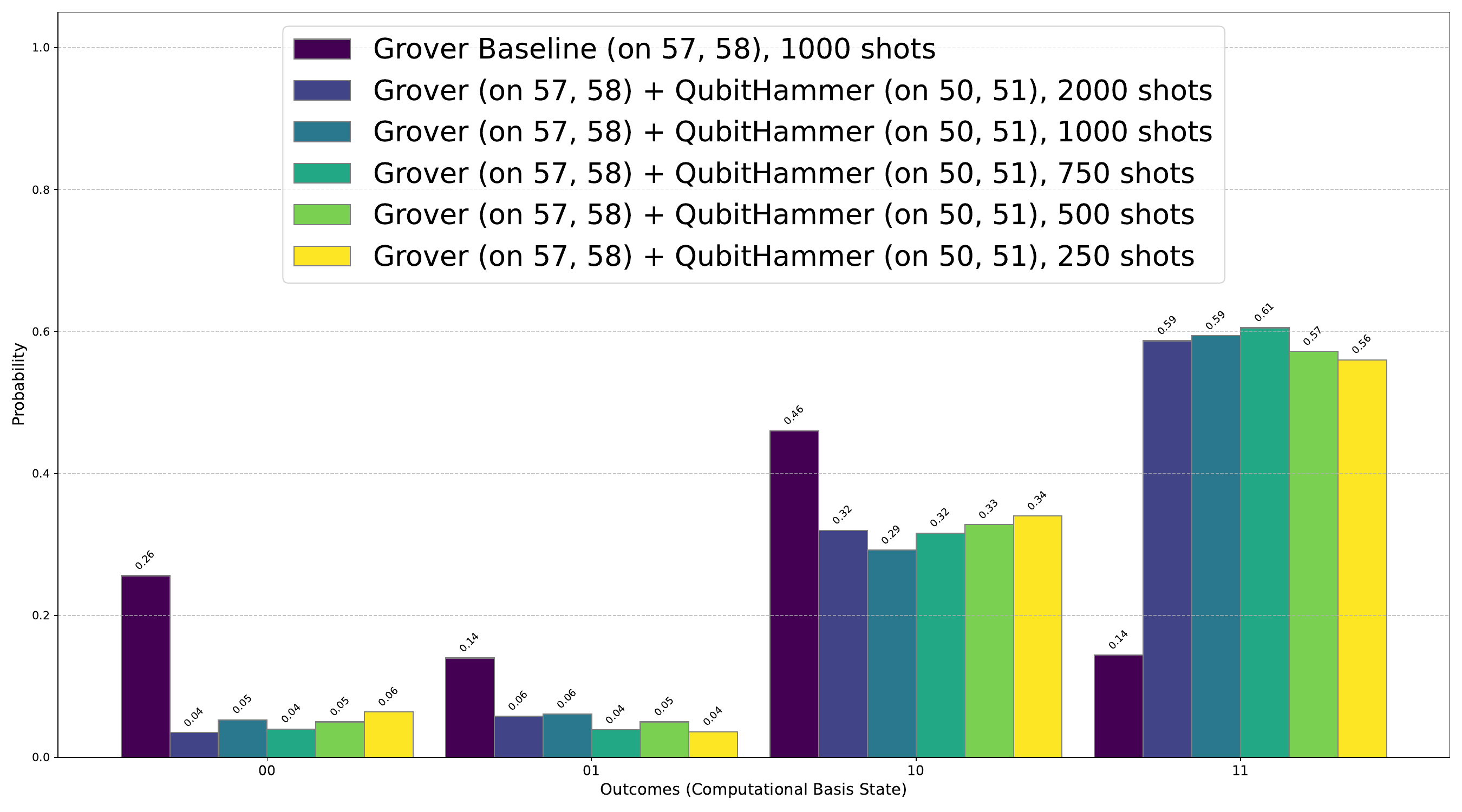}
 \caption{\small Evaluation of Attack Scenario 2 with a Repeated Pulses method using Grover circuit targeting qubit 57, 58 on \textit{Ankaa-3}.}
 \label{fig:rigetti_grover}
 \vspace{-3mm}
\end{figure}

To assess the impact on a practical application, we then targeted a 2-qubit Grover's algorithm running on victim qubits 57 and 58. Simultaneously, the adversary apply a 5-pulse {\tt X}-gate attack from two adjacent adversarial qubits 50 and 51.
The results, depicted in Figure~\ref{fig:rigetti_grover}, are striking. The repeated QubitHammer attack consistently forced the search algorithm to return an incorrect result \textit{`11'} instead of the correct solution \textit{`10'}, effectively corrupting the computation. This outcome is particularly noteworthy because the victim qubits (57, 58) were not among the most vulnerable identified in the full-chip scan, demonstrating the attack's potency even under non-ideal conditions.
Furthermore, we investigated the effect of reducing the number of measurement shots, which shortens the temporal overlap between the victim and attack circuits. The attack's success persisted even with fewer shots as shown in Figure~\ref{fig:rigetti_grover}. This demonstrates that prolonged interaction is not necessary to cause a critical failure, thereby alidating our third assumption regarding the adversary's capabilities in Section~\ref{subsec:assumptions}.

\subsection{Evaluation of Mixed Attack Pulse}

To demonstrate the efficacy and versatility of the mixed pulse attack, we conduct a comparative evaluation consisting of four distinct experiment on Rigetti \textit{Ankaa-3} processor. The victim was assigned qubits 57 and 58, while the adversary controlled qubits 50 and 51. The Bell state preparation circuit was selected due to its use of a sequence of native gates ({\tt RX($\pi$)}, {\tt RX($\pi$/2)}, {\tt iSWAP}), making it an ideal target to test the attack with mixed pulse structure.

For the baseline, the victim's two-qubit system was measured in an idle state without any adversarial interference to establish a baseline \textit{`00'} measurement.
%
Futher, the victim executed a Bell state preparation circuit to establish the baseline output 
distribution for a successful entanglement operation.
%
Next, we evaluated a conventional repeated-pulse attack, where the adversary applied 5 {\tt X}-gate attack pulses on his adversarial qubits while the victim's qubits were idle.
%
And finally, we deployed the mixed pulse attack. This involved the adversary synthesizing a malicious circuit that mirrored the victim's Bell state preparation, using native pulse parameters calibrated for the victim's qubits but applied to the adversary's own qubits. The victim qubits remain idle to ensure any resulting state change was due solely to the attack.

\begin{figure}[t]
\centering
\includegraphics[width=0.8\linewidth]{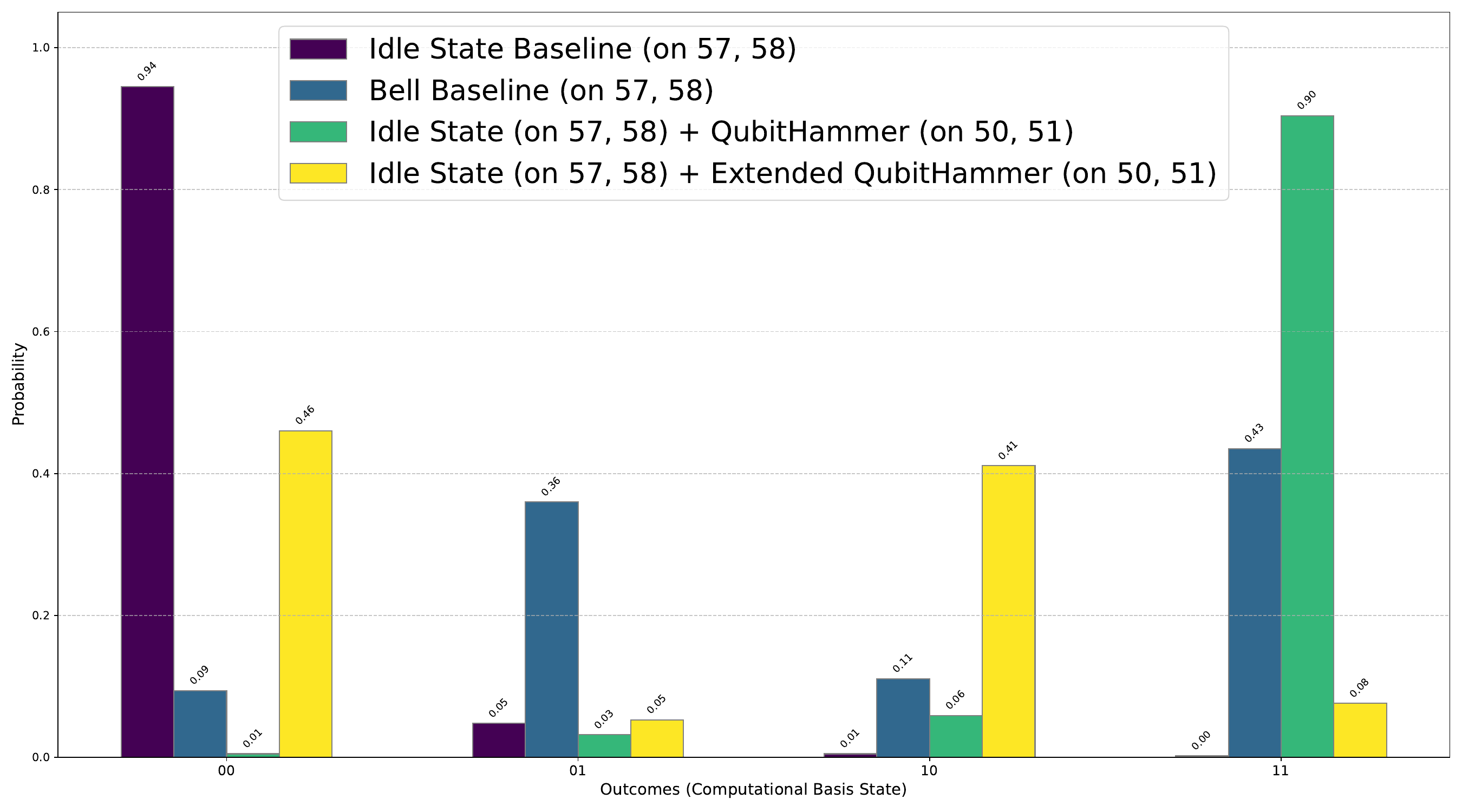}
 \caption{\small Evaluation of Extended QubitHammer attack with a Mixed Pulses method using Bell circuit targeting qubit 57, 58 on Rigetti \textit{Ankaa-3}.}
 \vspace{-4mm}
 \label{fig:rigetti_bell}
\end{figure}

The results, presented in Figure~\ref{fig:rigetti_bell}, reveal a critical difference in effectiveness between the attack methods, particularly in the type of error they induce. The conventional {\tt X}-gate attack produced a simple bit flip \textit{`00'} $\rightarrow$ \textit{`11'} with 0.90 probability, as expected. In stark contrast, the mixed pulse attack induced a more complex state corruption, preparing the idle victim in a near-equal superposition of \textit{`00'} and \textit{`10'} with a variational distance of 0.480. Moreover, the variational distance between the victim's Bell baseline and the mixed pulse attack reaches 0.668.
This outcome demonstrates that the mixed-pulse attack does not merely flip bits; it can actively steer the victim's idle qubits into a specific, non-trivial quantum state. This means that even without the victim running an explicit algorithm, the adversary can force the victim's qubits to prepare an arbitrary state, such as an unintended Bell state, simply by applying a carefully engineered pulse sequence.





%% file: sections/evaluation_summary.tex
\section{Overall Summary of QubitHammer}

\textcolor{black}{The physical mechanism behind QubitHammer is possibly unwanted resonant coupling, where driving an adversary's pulse at the victim's resonant frequency amplifies otherwise negligible parasitic interactions. While local crosstalk arises from direct capacitive coupling and imperfect qubit couplers, our long-range attacks likely exploit more subtle pathways. These include crosstalk between physically proximate control lines, signal propagation through chip substrate modes, and resonant modes within the chip's packaging. 
We demonstrated that this physical mechanisms behind QubitHammer affects major quantum computer hardware:
}

\begin{table*}[t]
\caption{\small Evaluation summary of QubitHammer targeting IBM quantum processors. The best (highest) variational distance for each attack scenario and method is summarized in this table. We assume that if the variational distance is larger than 0.2, then the corresponding attack is successful. The checkmark \Checkmark represents the success of the attack, while the cross \XSolidBrush means the attack failed.}
\vspace{-1mm}
\centering
\resizebox{0.95\linewidth}{!}{%
\renewcommand{\arraystretch}{1.5}
\begin{tabular}{c|cccccc}
\hline
\multirow{3}{*}{\textbf{Attack Scenario}} & \multicolumn{6}{c}{\textbf{Attack Method}}                                          \\ \cline{2-7} 
                         & \multicolumn{2}{c|}{\textit{Single Attack Pulse on Single Victim Qubit}} & \multicolumn{2}{c|}{\textit{Repeated Attack Pulses on Grover}} & \multicolumn{2}{c}{\textit{Repeated Attack Pulses on QAOA}} \\ \cline{2-7}
                         & \multicolumn{1}{c|}{\textit{Variational Distance}} & \multicolumn{1}{c|}{\textit{Attack Succeeded}} & \multicolumn{1}{c|}{\textit{Variational Distance}} & \multicolumn{1}{c|}{\textit{Attack Succeeded}} & \multicolumn{1}{c|}{\textit{Variational Distance}} & \textit{Attack Succeeded} \\ \hline
\textit{Scenario 1 (BAI)} & \multicolumn{1}{c|}{0.609} & \multicolumn{1}{c|}{\Checkmark} & \multicolumn{1}{c|}{0.801} & \multicolumn{1}{c|}{\Checkmark} & \multicolumn{1}{c|}{-} & - \\ \hline
\textit{Scenario 1 (CAI)} & \multicolumn{1}{c|}{0.750} & \multicolumn{1}{c|}{\Checkmark} & \multicolumn{1}{c|}{0.578} & \multicolumn{1}{c|}{\Checkmark} &\multicolumn{1}{c|}{0.17} & \XSolidBrush \\ \hline
\textit{Scenario 2 (BAI)} & \multicolumn{1}{c|}{0.753} & \multicolumn{1}{c|}{\Checkmark} & \multicolumn{1}{c|}{0.770} & \multicolumn{1}{c|}{\Checkmark} &  \multicolumn{1}{c|}{0.307} & \Checkmark \\ \hline
\textit{Scenario 2 (CAI)} & \multicolumn{1}{c|}{0.688} & \multicolumn{1}{c|}{\Checkmark} & \multicolumn{1}{c|}{0.664} & \multicolumn{1}{c|}{\Checkmark} & \multicolumn{1}{c|}{-} & - \\ \hline
\end{tabular}%
}
\label{tab:ibm_eval_summary}
\end{table*}

{\textbf{IBM Quantum Processors:}}
Table~\ref{tab:ibm_eval_summary} summarizes the effectiveness of the attacks on IBM quantum processors. The first column summarizes the attack scenario utilized. Columns 2, 4, and 6 denote the maximal variational distance $D_{TV}$ obtained by our experiments for three attack methods, respectively. Columns 3, 5, and 7 show whether the attack is successful or not. QubitHammer operating with Attack Scenario~1 is most potent ($D_{TV}$ up to \(0.801\) with repeated pulses) but less practical due to its large adversarial footprint, while the more realistic Scenario~2 (CAI) still reaches \(D_{TV} = 0.688\) (single pulse) and \(D_{TV} = 0.664\) (repeated). 
Even under compiler optimization level~3, we observe \(D_{TV}\) as high as \(0.938\). 
Our evaluations discovered that qubit 0 on three IBM machines, \textit{IBM\_brisbane}, \textit{IBM\_kyoto}, and \textit{IBM\_osaka}, is extremely vulnerable to the QubitHammer attacks. Consequently, any circuit using qubit 0 can be manipulated easily.

{\textbf{Rigetti Quantum Processors:}}
Table~\ref{tab:rigetti_eval_summary} summarizes the evaluation on the \textit{Rigetti} \textit{Ankaa-3} processor, distinguishing between the attack's impact (the type of error induced) and its overall effectiveness, as quantified by the variational distance. While single attack pulse method proved insufficient to reliably corrupt a victim qubit ($D_{TV}$ = 0.182), the repeated attack pulse method was highly effective ($D_{TV}$ = 0.558), confirming a successful deployment of QubitHammer. The threat profile on Rigetti differs from that on the IBM system: although the peak impact on any single qubit is lower, the vulnerability is more widespread across the chip. Furthermore, the success of the mixed attack pulse method shows a more advanced threat, as its ability to cause complex errors in the quantum state, instead of just simple bit-flips, highlights the flexibility of our QubitHammer attacks.

\begin{table}[t]
\centering
\caption{\small Overall summary of the attacks and their effectiveness targeting Rigetti \textit{Ankaa-3} processor.}
\vspace{-1mm}
\renewcommand{\arraystretch}{2} 
\setlength{\tabcolsep}{3pt} 
\resizebox{\linewidth}{!}{%
\begin{tabular}{c|c|c|c}
\hline
\textbf{Attack Method}    &\textbf{Variational Distance}   & \textbf{Attack Impact}  & \textbf{Attack Succeeded} \\ \hline
\textit{Single Attack Pulse} & 0.182   & Qubit flipping  & \XSolidBrush   \\ \hline
\textit{Repeated Attack Pulse} & 0.558   & Qubit flipping  & \Checkmark   \\ \hline
\textit{Mixed Attack Pulse} & 0.480  & Complex state corruptions  & \Checkmark   \\ \hline
\end{tabular}%
}
\label{tab:rigetti_eval_summary}
\end{table}

%% file: sections/defenses.tex
\section{Bypassing Existing Protections}


We now show, in Table~\ref{tab:eval_summary_defeneses}, that every current crosstalk defense can be bypassed by at least one of our attack scenarios.

\begin{figure}[t]
\centering
\begin{subfigure}[b]{0.49\linewidth}
\includegraphics[width=\textwidth]{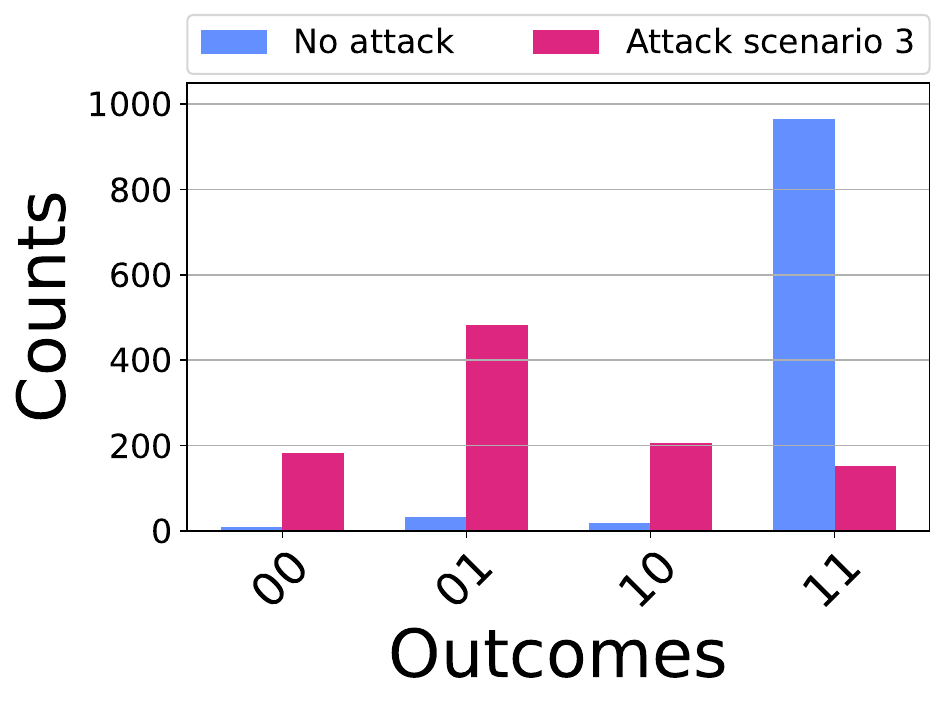}
\caption{\small IBM\_brisbane}
\label{subfig:res_dd_1}
\end{subfigure}
\begin{subfigure}[b]{0.49\linewidth}
\includegraphics[width=\textwidth]{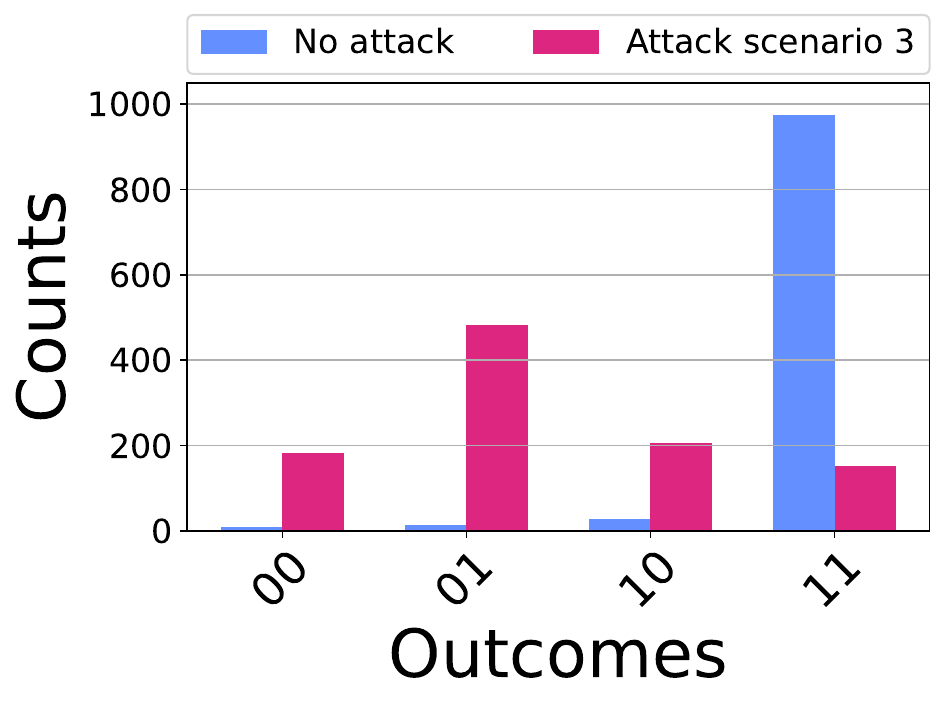}
\caption{\small IBM\_kyoto}
\label{subfig:res_dd_2}
\end{subfigure}
\vspace{-6mm}
\caption{\small Evaluation of Attack Scenario 2 (BAI) using dynamical decoupling on \textit{IBM\_brisbane} and \textit{IBM\_kyoto}.}
\label{fig:dd_compiled_res}
\end{figure}

\textcolor{black}{
\noindent\textbf{Dynamical Decoupling.} This well-established error suppression method inserts pulse sequences on idle qubits to reduce interqubit crosstalk errors~\cite{bylander2011noise}~\cite{tripathi2022suppression}. 
Our five-pulse Scenario 2 (BAI) attack against a Grover's circuit protected by dynamical decoupling yielded high variational distances of 0.747 on \textit{IBM\_brisbane} and 0.837 on \textit{IBM\_kyoto} in Figure~\ref{fig:dd_compiled_res}.}
Other attack scenarios also demonstrated similar degradation in performance confirming that dynamical decoupling fails to defend against QubitHammer.

\begin{figure}[t]
\centering
\includegraphics[width=0.55\linewidth]{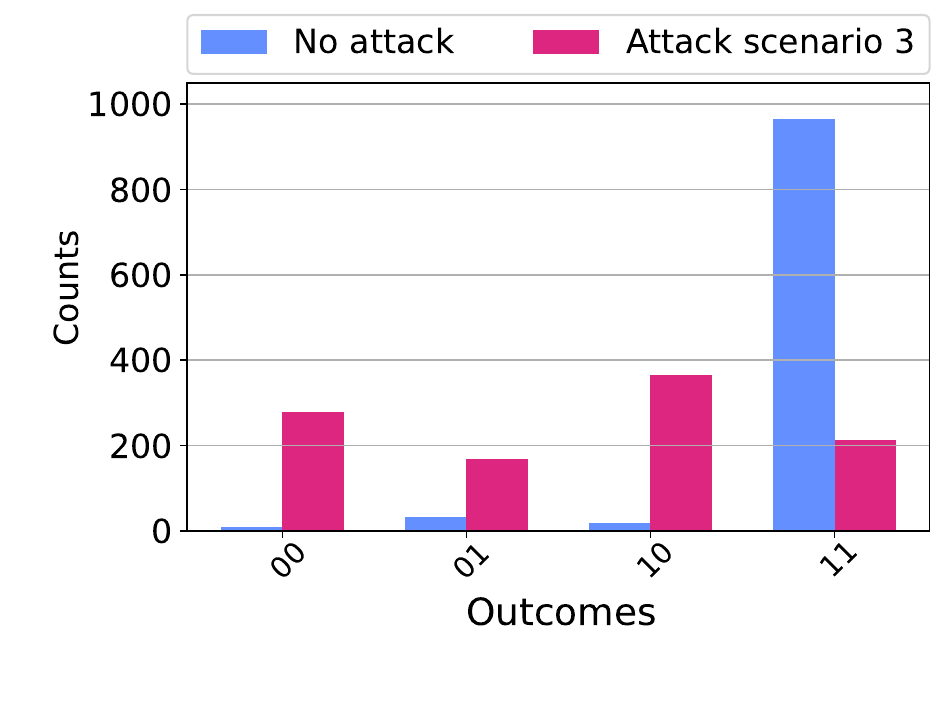}
\vspace{-2mm}
\caption{\small Evaluation of Attack Scenario 2 (BAI) with padding on idle qubits on \textit{IBM\_brisbane}.}
\label{fig:pad_defense}
\end{figure}

\begin{table}[t]
\caption{\small Effectiveness of QubitHammer attacks against existing countermeasures.}
\vspace{-1mm}
\centering
\resizebox{\linewidth}{!}{%
\renewcommand{\arraystretch}{1.2}
\begin{tabular}{c|c|c}
\hline
\textbf{Defense}        & \textbf{Defense Description}                                                                                & \textbf{Attack Success} \\ \hline
Dynamical decoupling    & \begin{tabular}[c]{@{}c@{}}Add dynamical decoupling \\ sequences to program\end{tabular}                    & \Checkmark              \\ \hline
Disabling highly vulnerable qubit & \begin{tabular}[c]{@{}c@{}}Removing highly vulnerable \\ qubit during qubit allocation\end{tabular}       & \Checkmark \\ \hline
Crosstalk-aware qubit allocation  & \begin{tabular}[c]{@{}c@{}}Allocating idle qubits as padding \\ between two quantum circuits\end{tabular} & \Checkmark \\ \hline
Active padding          & \begin{tabular}[c]{@{}c@{}}Allocating active qubits as padding \\ between two quantum circuits\end{tabular} & \Checkmark              \\ \hline
Disabling custom pulses & \begin{tabular}[c]{@{}c@{}}Disabling deployment \\ of user-designed custom pulses\end{tabular}              & \textbf{---}            \\ \hline
\end{tabular}%
}
\label{tab:eval_summary_defeneses}
\end{table}

\textbf{Crosstalk-aware qubit allocation.} This approach allocates idle qubits as padding between different quantum programs to minimize the effect of crosstalk~\cite{share}. However, as explained in Section~\ref{subsec:atk_3}, such a defense is ineffective against our proposed~attack.

\textcolor{black}{
\noindent\textbf{Active padding.} 
We also evaluated active padding, a defense where buffer qubits are driven with identity-equivalent gate sequences (e.g. an even number of X-, XY-, or Y-gates) to counteract crosstalk from idle qubits.
Despite this defense, our Scenario 2 (BAI) attack against a Grover's circuit remained highly effective as shown in Figure~\ref{fig:pad_defense}, yielding a variational distance of 0.73. This result demonstrates QubitHammer can bypass defenses that rely on active noise suppression.}

\textcolor{black}{
\noindent\textbf{Disabling highly vulnerable qubit.} A naive hardware fix, such as disabling highly vulnerable qubits, is an insufficient defense. This approach merely conceals the underlying hardware susceptibility, which our cross-platform analysis shows is not an isolated issue.
}

\textcolor{black}{
\noindent\textbf{Disabling custom pulses.} A final line of defense could be to detect and block malicious pulse sequences. However, distinguishing a malicious QubitHammer pulse sequence from a legitimate, complex pulse sequence used in applications like quantum machine learning is an open problem. Further, custom pulses have valid applications in quantum machine learning, so disabling such functionality is impractical.}


%% file: sections/related_work.tex
\section{Related Work}



Various existing multi-tenant schemes have been proposed to optimize throughput by allowing multiple users to run their circuits on the same quantum computer simultaneously but utilizing different qubits~\cite{das2019case}~\cite{liu2021qucloud}. While this concept holds significant promise for enhancing the utilization of quantum computing resources, it also exposes quantum computers to security vulnerabilities caused by crosstalk errors~\cite{sarovar2020detecting}~\cite{murali2020software}~\cite{deshpande2022towards}. 

Malicious users can exploit crosstalk in various ways. For instance, {\tt CNOT} gates can induce significant crosstalk, allowing attackers to design circuits that disrupt nearby qubits instead of performing valid computations~\cite{harper2024crosstalk}. If aware of QPU co-tenancy, attackers may also extract sensitive information from state-preparation circuits~\cite{harper2024crosstalk}. Furthermore, crosstalk can be used to target specific quantum algorithms, such as ensuring that Grover’s algorithm consistently returns incorrect results, thereby sabotaging more complex quantum jobs~\cite{deshpande2022towards}. Unlike our work, none of these consider custom control pulses, nor do they work when victim and attacker are far away.

%% file: sections/conclusion.tex
\section{Conclusion}

\balance

In this work, we introduced QubitHammer, a novel set of qubit-flipping attacks targeting state-of-the-art superconducting quantum computers. Our attacks were validated on quantum computers from IBM and Rigetti. The results highlighted significant security and reliability concerns. Further, the attacks were demonstrated to bypass  existing defenses that have been so-far proposed for defending against crosstalk-based attacks, and thus we demonstrated a need to develop new types of defenses before safe and secure multi-tenant quantum computers can be achieved.

%% file: sections/ethical_consideration.tex
\section{Ethical Considerations}

This work focuses on evaluation of novel threats in quantum computers. While multi-tenancy is widely proposed, it is not yet implemented by cloud-based quantum computer providers. As result, this study provides necessary insights about dangers of shared quantum computers, but poses no ethical risk: we are highlighting a new problem before it may widely affect quantum computers giving the community time to understand it and develop mitigations.